\documentclass[journal]{IEEEtranTIE}
\usepackage{graphicx}
\usepackage{cite}
\usepackage{picinpar}
\usepackage{amsmath}
\usepackage{url}
\usepackage{flushend}
\usepackage[latin1]{inputenc}
\usepackage{colortbl}
\usepackage{soul}
\usepackage{multirow}
\usepackage{pifont}
\usepackage{color}
\usepackage{alltt}
\usepackage[hidelinks]{hyperref}
\usepackage{enumerate}
\usepackage{siunitx}
\usepackage{breakurl}
\usepackage{epstopdf}
\usepackage{pbox}
\usepackage{amsfonts}
\usepackage{bm}
\usepackage{subcaption}
\usepackage{amssymb}
\usepackage{enumitem}

\newtheorem{theorem}{Theorem}
\newtheorem{assumption}{Assumption}
\newtheorem{lemma}{Lemma}
\newtheorem{remark}{Remark}
\newtheorem{property}{Property}

\begin{document}
\title{Adaptive Task Space Non-Singular Terminal Super-Twisting Sliding Mode Control of a 7-DOF Robotic Manipulator}

\author{
	\vskip 1em
	
	Lucas Wan, \emph{Student Member, IEEE},
    Sean Smith, \emph{Student Member, IEEE},
    Ya-Jun Pan, \emph{Senior Member, IEEE},
    Emmanuel Witrant

	\thanks{
This work was supported by the Natural Sciences and Engineering Research Council (NSERC) and the Government of Nova Scotia, Canada.
Lucas Wan and Ya-Jun Pan are with the Department of Mechanical Engineering, Dalhousie University, Halifax NS B3H 4R2, Canada (e-mail: lucas.wan@dal.ca; yajun.pan@dal.ca). Sean Smith and Emmanuel Witrant are with the GIPSA-lab, CNRS, University of Grenoble Alpes, F-38000 Grenoble, France and the Department of Mechanical Engineering, Dalhousie University, Halifax NS B3H 4R2, Canada (e-mail: s.smith@dal.ca; emmanuel.witrant@univ-grenoble-alpes.fr).
	}
}

\maketitle

\IEEEpubid{\begin{minipage}{\textwidth}\centering
\vspace{6\baselineskip}
\footnotesize
\copyright 2025 IEEE. Personal use of this material is permitted. 
Permission from IEEE must be obtained for all other uses, in any current or future media, including reprinting/republishing this material for advertising or promotional purposes, creating new collective works, for resale or redistribution to servers or lists, or reuse of any copyrighted component of this work in other works. This is the author's version of an article that has been accepted for publication in IEEE Transactions on Industrial Electronics. 
The final version is available at: \url{https://doi.org/10.1109/TIE.2025.3600520}
\end{minipage}}
	
\begin{abstract}
This paper presents a new task-space Non-singular Terminal Super-Twisting Sliding Mode (NT-STSM) controller with adaptive gains for robust trajectory tracking of a 7-DOF robotic manipulator. The proposed approach addresses the challenges of chattering, unknown disturbances, and rotational motion tracking, making it suited for high-DOF manipulators in dexterous manipulation tasks. A rigorous boundedness proof is provided, offering gain selection guidelines for practical implementation. Simulations and hardware experiments with external disturbances demonstrate the proposed controller's robust, accurate tracking with reduced control effort under unknown disturbances compared to other NT-STSM and conventional controllers. The results demonstrated that the proposed NT-STSM controller mitigates chattering and instability in complex motions, making it a viable solution for dexterous robotic manipulations and various industrial applications.
\end{abstract}


\markboth{IEEE TRANSACTIONS ON INDUSTRIAL ELECTRONICS}%
{}

\definecolor{limegreen}{rgb}{0.2, 0.8, 0.2}
\definecolor{forestgreen}{rgb}{0.13, 0.55, 0.13}
\definecolor{greenhtml}{rgb}{0.0, 0.5, 0.0}

\section{Introduction}


\IEEEPARstart{T}{he} development of robust control algorithms is necessary for industrial robotic manipulators in applications such as remote surgery, cooperative multi-robot manipulation, and handling varying payloads. These applications require precise trajectory tracking, robustness to disturbances, and energy-efficient control strategies. High degree-of-freedom (DOF) manipulators offer an extensive range of motion, however, their complex nonlinear dynamics, with model uncertainties and external disturbances, pose significant control challenges. 

Sliding mode (SM) control provides precise tracking capabilities and is robust against disturbances and uncertainties. Non-singular Terminal Sliding Mode (NTSM) control offers fast, finite-time convergence speed and avoids singularities, making it appealing for robotic manipulators \cite{feng2002non}. However, its high-frequency switching causes undesirable chattering and high control effort, which risks exceeding actuator limits, increased energy usage, and excessive mechanical wear, limiting its practical implementation. Super-Twisting Sliding Mode (STSM) control \cite{moreno2012strict} alleviates this issue by smoothing the control signal, making it suitable for robotic applications \cite{jeong2018tracking}. 

In recent years, SM control has been widely applied to joint space manipulator control. Fixed-gain approaches include fractional-order NTSM controllers \cite{wang2016practical} and fixed-time terminal SM controllers \cite{sai2022approximate}, which provide fast convergence and accurate trajectory tracking. Adaptive gain strategies further improve tracking performance, reduce control effort, and relax the need for prior disturbance bounds \cite{shtessel2012novel, gonzalez2011variable}. Adaptive SM control has been combined with time-delay estimation \cite{baek2016new}, fault-tolerant control \cite{zhang2021adaptive}, and neural-network uncertainty modeling \cite{sun2022adaptive, hu2022robust}. However, these controllers are complex, remain limited to low-DOF manipulators (2 to 4-DOF), and are often applied only in simulation.

One exception is \cite{jie2020trajectory}, which applied a fractional-order terminal SM controller to a real 7-DOF manipulator. Yet this work only validates simple joint trajectories and does not include external disturbances or payload variation. Another disadvantage of these works is that joint space controllers require converting task space objectives to joint references, introducing susceptibility to modeling errors and uncertainties \cite{veysi2015novel}. This motivates task-space control approaches in certain applications requiring precise Cartesian tracking.

Task space control, where reference trajectories are defined in Cartesian space and used directly in the controller, is effective for robotic tasks such as high-precision assembly, force interactions, and collaborative manipulation \cite{lewis2003robot}. However, robust task space SM control remains less explored, especially for high-DOF systems, due to challenges in handling both translational and rotational motion simultaneously. Existing task-space SM control approaches are limited to low-DOF manipulators \cite{veysi2015novel}, or focus only on translational motion \cite{khan2011task, nicolis2020operational}. A recent task-space adaptive SM controller was developed for a 4-DOF Delta robot \cite{fateh2024model}, but it is limited to simulation. Moreover, few works consider chattering in rotational motion, which is sensitive to this problem and requires consideration of rotational representations.

The combination of NTSM and STSM control into a non-singular terminal super-twisting SM (NT-STSM) controller has been explored in \cite{yi2019adaptive, zhai2021fast, cruz2022non, sun2024adaptive}. However, existing controllers are designed in joint space, validated only on low-DOF platforms, and often impose restrictive gain relationships in their stability proofs. This lack of experimental validation and flexibility highlights the need for a practical task-space NT-STSM design applicable to high-DOF manipulators.

Despite the theoretical advantages of SM control, its practical use in real-time, high-DOF robot applications remains limited due to chattering, challenges in gain tuning, and a lack of experimental validation. Existing NT-STSM designs focus on joint space and do not address the unique challenges of tracking rotational motion in Cartesian space. This paper addresses these gaps by proposing a task-space NT-STSM controller with adaptive gains guaranteeing convergence and robustness through a new stability analysis. The controller is experimentally validated on a 7-DOF manipulator in a task with a time-varying payload, demonstrating practical feasibility and improved performance. In summary, the proposed approach applies to the control of robotic manipulators with an arbitrary number of DOFs for dexterous manipulations. The contributions of this work are:


\begin{enumerate}[leftmargin=*]
    \item An NT-STSM controller is proposed with an adaptive gain law designed to improve tracking performance and reduce chattering without requiring prior knowledge of disturbance bounds. A novel boundedness proof is derived with tunable gain selection guidelines for practical implementation. Compared to \cite{sun2024adaptive}, the proposed boundedness proof relaxes the relationship between the STSM gains, providing improved performance on a 7-DOF robotic manipulator.
    \item  The proposed NT-STSM controller is designed in task space with unit quaternions representing orientation. Unlike prior task-space SM control approaches \cite{veysi2015novel, khan2011task, nicolis2020operational}, this work explicitly incorporates unit quaternions, allowing for efficient rotational motion while preventing kinematic singularities. As demonstrated in Section \ref{sec:results}, the rotational motion can be more prone to chattering, reinforcing the importance of the proposed chatter-free task space NT-STSM controller.
    \item The proposed NT-STSM controller's tracking performance and robustness against unknown disturbances are validated with simulations and experiments on a 7-DOF Franka Emika Robot with extensive comparisons. The proposed controller demonstrates lower position and orientation tracking error while maintaining low control effort and low chatter, compared to the conventional PD control and NT-STSM controller developed in \cite{sun2024adaptive}. Unlike NTSM and STSM controllers, which suffer from severe chattering and instability, the proposed controller maintains smooth trajectory tracking.
\end{enumerate}

\section{Preliminaries} 

\subsection{Dynamic Model} \label{sec:dynamicmodel}
Consider a $7$-DOF robot manipulator with the joint space dynamics of the form \cite[Ch. 3]{siciliano2008springer}
	\begin{equation} \label{eqELjs}
	\mathcal{M}(\bm{q})\ddot{\bm{q}}+\mathcal{C}(\bm{q},\dot{\bm{q}})\dot{\bm{q}} + \mathcal{G}(\bm{q}) = \bm{\tau} + \bm{\tau}_{d}(\bm{q}),
	\end{equation}
	where $\bm{q}, \dot{\bm{q}}, \ddot{\bm{q}} \in \mathbb{R}^{7}$ are the joint position, velocity, and acceleration vectors, respectively, $\mathcal{M}(\bm{q}) \in \mathbb{R}^{7 \times 7}$ is the inertia matrix, $\mathcal{C}(\bm{q},\dot{\bm{q}}) \in \mathbb{R}^{7 \times 7}$ is the Coriolis and centripetal torque matrix, $\mathcal{G}(\bm{q}) \in \mathbb{R}^{7}$ is the gravitational torque vector, $\bm{\tau} \in \mathbb{R}^{7}$ is the joint torque control input vector, $\bm{\tau}_{d}(\bm{q}) = \bm{\tau}_{f}(\bm{q}) + \bm{\tau}_{u}(\bm{q}, \dot{\bm{q}}) + \bm{\tau}_{e} \in \mathbb{R}^{7}$ is the summation of the internal torque friction, $\bm{\tau}_{f}(\bm{q})$, the unmodelled dynamics due to modelling inaccuracies, $\bm{\tau}_{u}(\bm{q}, \dot{\bm{q}})$, and the external torque exerted on the manipulator, $\bm{\tau}_{e}$. In task space, the dynamics of the robot manipulators may be represented as \cite[Ch. 3]{siciliano2008springer}
	\begin{equation} \label{eqELts}
	\bar{\mathcal{M}}(\bm{q})\ddot{\bm{x}}+\bar{\mathcal{C}}(\bm{q},\dot{\bm{q}}) + \bar{\mathcal{G}}(\bm{q}) = \bm{u} + \bm{f}_{d}(\bm{q}).
	\end{equation}
The end-effector pose in task space is denoted as $\bm{x} = \begin{bmatrix}
\bm{p}^T, & \bm{\xi}^T
\end{bmatrix}^T \in \mathbb{R}^{7}$ where $\bm{p} \in \mathbb{R}^{3}$ is the translational position and $\bm{\xi} = \begin{bmatrix}
\eta, & \bm{\epsilon}^T
\end{bmatrix}^T \in \mathrm{S}^{3} \subset \mathbb{R}^4$ is the orientation in the form of a unit quaternion. The quaternion is made up of $\eta$, a scalar denoting the real part, and $\bm{\epsilon} = \begin{bmatrix}
\epsilon_x, & \epsilon_y, & \epsilon_z
\end{bmatrix}^T$, a vector denoting the imaginary part. The end-effector velocity and acceleration are denoted as $\dot{\bm{x}} = \begin{bmatrix}
\dot{\bm{p}}^T, & \bm{\omega}^T
\end{bmatrix}^T \in \mathbb{R}^{6}$ and $\ddot{\bm{x}} = \begin{bmatrix}
\ddot{\bm{p}}^T, & \dot{\bm{\omega}}^T
\end{bmatrix}^T \in \mathbb{R}^{6}$, where $\bm{\omega}, \ \dot{\bm{\omega}} \in \mathbb{R}^3$ are the angular velocity and acceleration of the end-effector, respectively.


	
The end-effector velocity and joint velocity are related by the Jacobian matrix, as $\dot{\bm{x}} = J(\bm{q})\dot{\bm{q}}$. For the remainder of the paper, the dependencies of the dynamic terms on $\bm{q}$ and $\dot{\bm{q}}$ will be omitted. 

	%
	\begin{property}
	The Cartesian parametric matrices in \eqref{eqELts} are related to the joint space matrices in \eqref{eqELjs}
by \cite[Ch. 3]{siciliano2008springer}
\begin{align*}
\bm{\tau} &= J^T \bm{u}, &&\bar{\mathcal{M}} = (J \mathcal{M}^{-1} J^T)^{-1}, \\
\bar{\mathcal{C}} &= \bar{\mathcal{M}} (J \mathcal{M}^{-1} \mathcal{C} \dot{\bm{q}} - \dot{J} \dot{\bm{q}}), &&\bar{\mathcal{G}} = \bar{\mathcal{M}} J \mathcal{M}^{-1} \mathcal{G}. 
\end{align*}
	\end{property}
	\begin{property}
	The inertia matrix, $\mathcal{M}$, is positive-definite symmetric, $\dot{\mathcal{M}} - 2\mathcal{C}$ is skew-symmetric, and the Jacobian matrix, $J$, has full rank \cite[Ch. 3]{siciliano2008springer}.
	\end{property}

\subsection{Trajectory Generation} \label{sec:traj}
The trajectory generator provides the controller with a desired trajectory, $\begin{bmatrix}
\bm{x}_d, & \dot{\bm{x}}_d, & \ddot{\bm{x}}_d
\end{bmatrix}^T$. This trajectory is interpolated between an initial pose, $\bm{x}_{d_0}$, and goal pose, $\bm{x}_{d_g}$.

The translational motion is calculated using the clamped cubic spline approach to provide smooth trajectories with zero velocities at the start and end. Unit quaternions represent the orientation position, therefore, standard interpolation methods may not result in consistent and smooth rotational trajectories \cite{mukundan2012advanced}. To interpolate between an initial quaternion, $\bm{\xi}_{d_0}$, and a goal quaternion $\bm{\xi}_{d_g}$, the cubic Hermite curve interpolation method \cite{kreyszig2008advanced} is used: 
\begin{align} \label{eq:desquat}
	\bm{\xi}_d(t) &= \frac{F_{a2q}(W F_{q2a}(\bm{\xi}_{d_g} \times \bm{\xi}_{d_0}^{-1}) \times \bm{\xi}_{d_0})}{|F_{a2q}(W F_{q2a}(\bm{\xi}_{d_g} \times \bm{\xi}_{d_0}^{-1}) \times \bm{\xi}_{d_0})|}, \\
	W &= 3 \left( \frac{t}{T} \right)^2 - 2 \left( \frac{t}{T} \right)^3, \nonumber
\end{align}
where $T$ is the time between two poses, $\bm{\xi}_1 \times \bm{\xi}_2$ represents quaternion multiplication, $|\bm{\xi}|$ represents the norm of the quaternion, and $\bm{\xi}^{-1}$ represents the inverse of the quaternion, which in the case of unit quaternions is equal to the conjugate of the quaternion, $\bm{\xi}^*$. $F_{q2a}(\bm{\xi})$ represents a function to convert a quaternion to a scaled angle axis, $\bm{\Theta}$, and $F_{a2q}(\bm{\Theta})$ represents a function to convert a scaled angle axis to a quaternion. 

The desired angular velocity is calculated from \eqref{eq:desquat} using the quaternions as \cite{kou2018linear}
\begin{equation}
	\bm{\omega}_d(t) = \frac{2}{\Delta t} \begin{bmatrix}
	 \eta' \epsilon_x - \epsilon_x' \eta - \epsilon_y' \epsilon_z + \epsilon_z' \epsilon_y \\
	-\eta' \epsilon_y - \epsilon_x' \epsilon_z + \epsilon_y' \eta + \epsilon_z' \epsilon_x \\
	- \eta' \epsilon_z + \epsilon_x' \epsilon_y - \epsilon_y' \epsilon_x + \epsilon_z' \eta
	\end{bmatrix},
\end{equation}
where $\Delta t$ is the sampling time, $\eta = \eta_d(t)$, $\bm{\epsilon} = \bm{\epsilon}_d(t)$, $\eta' = \eta_d(t-\Delta t)$, and $\bm{\epsilon}' = \bm{\epsilon}_d(t - \Delta t)$. The desired angular acceleration is calculated using the finite difference method.

\subsection{Third Order Sliding Mode Observer}
%
%
SM observers provide a robust method of estimating the velocity of robotic manipulators. According to experimental tests in \cite{liu2022velocity}, SM observers exhibit consistent and high accuracy over a range of sensor resolutions and sampling rates, whereas moving average and derivative filters often had significant estimation errors. SM observers have the additional benefit of being robust to inconsistent sensor readings \cite{liu2022velocity}. Similar to SM controllers, SM observers suffer from chattering in real-world applications. An approach to reducing chatter is to use a TOSM observer, although they have slower convergence than lower-order observers. In \cite{van2013output}, linear terms are introduced to increase the convergence speed. By assigning $\bm{g}_1 = \bm{x}$, $\bm{g}_2 = \dot{\bm{x}}$, and denoting $\hat{\bm{g}}_1$ and $\hat{\bm{g}}_2$ as the estimated states, a TOSM observer is adapted from \cite{van2013output} and formulated in task space
\begin{align}
	\dot{\hat{\bm{g}}}_1 =& \ \hat{\bm{g}}_2 + \alpha_{o2} \odot |\bm{g}_1 - \hat{\bm{g}}_1|^{\frac{2}{3}} \odot \textrm{sign} (\bm{g}_1 - \hat{\bm{g}}_1) \nonumber \\
    & \hspace{-1mm} + k_{o2} (\bm{g}_1 - \hat{\bm{g}}_1), \label{eq:ob1} \\
	\dot{\hat{\bm{g}}}_2 =& \ \bar{\mathcal{M}}^{-1} (\bm{u} - \bar{\mathcal{C}} - \bar{\mathcal{G}}) + \alpha_{o1} \odot |\dot{\hat{\bm{g}}}_1 - \hat{\bm{g}}_2|^{\frac{1}{2}} \odot \textrm{sign} (\dot{\hat{\bm{g}}}_1 - \hat{\bm{g}}_2) \nonumber \\ 
	& \hspace{-1mm} + k_{o1} (\bm{g}_1 - \hat{\bm{g}}_1) + \hat{\bm{z}}, \label{eq:ob2} \\
	\dot{\hat{\bm{z}}} =& \ \alpha_{o0} \textrm{sign} (\dot{\hat{\bm{g}}}_1 -  \hat{\bm{g}}_2), \label{eq:ob3}
\end{align}
where $\alpha_{o2}$, $\alpha_{o1}$, $\alpha_{o0} \in \mathbb{R}^{7}$ are positive gain vectors, $k_{o2}$, $k_{o1}$ are positive scalar gains, and $\odot$ is the Hadamard product. The estimation error of the observed states are defined as $\tilde{\bm{g}}_1 = \bm{g}_1 - \hat{\bm{g}}_1$, and $\tilde{\bm{g}}_2 = \bm{g}_2 - \hat{\bm{g}}_2$, detailed in \cite[Eq. (29)]{van2013output}. The convergence proof for the TOSM observer is in \cite[Theorem 1]{van2013output}, showing that $\tilde{\bm{g}}_1$ and $\tilde{\bm{g}}_2$ are stable and converge to zero in finite time. In implementation, first-order approximation methods are used to integrate the observer terms.




Despite introducing the linear terms to the observer to increase the convergence speed, there is still an initial period during which the estimated states are inaccurate before they converge. This can lead to unpredictable behaviour of the manipulator. A finite difference method and an exponential moving average filter are used to estimate the velocity during this convergence period as
\begin{equation*}
	\dot{\bar{\bm{x}}}(t) = \alpha_e \frac{\bm{x}(t) - \bm{x}(t - \Delta t)}{\Delta t} + (1 - \alpha_e) \dot{\bar{\bm{x}}}(t - \Delta t),
\end{equation*}
where $\alpha_e$ is a positive filter design parameter.
The position and velocity sent to the controller are selected as
\begin{align}
    \hat{\bm{x}} &=
	\begin{cases}
		\bm{x} & \text{if } | \hat{\bm{g}}_1 - \bm{x} | > \eta_q \\
		\hat{\bm{g}}_1 & \text{if } | \hat{\bm{g}}_1 - \bm{x} | \leq \eta_q
	\end{cases}, \\
	\dot{\hat{\bm{x}}} &=
	\begin{cases}
		\dot{\bar{\bm{x}}} & \text{if } | \hat{\bm{g}}_2 - \dot{\bar{\bm{x}}}| > \eta_q \\
		\hat{\bm{g}}_2 & \text{if } | \hat{\bm{g}}_2 - \dot{\bar{\bm{x}}}| \leq \eta_q
	\end{cases},
\end{align}
where $\eta_q > 0$ is a positive design parameter.

\section{Adaptive Robust Controller Design} \label{sec:controller}

\begin{figure}
\centering
\includegraphics[width=\linewidth]{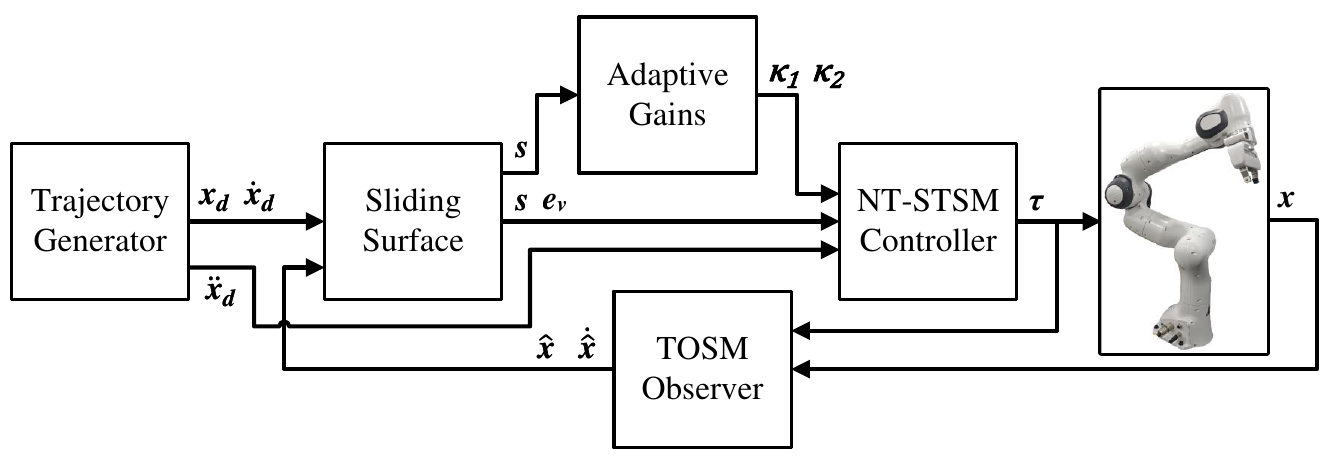}
\caption{Block diagram of the proposed control design.}
\label{fig:block}
\end{figure}

In this section, a novel NT-STSM controller is proposed along with a rigorous boundedness proof. The block diagram of the controller and observer is shown in Fig. \ref{fig:block}.
\subsection{Non-Singular Terminal Super-Twisting Sliding Mode Controller}
%
%
Compared to the conventional SM control, an NTSM controller offers appealing characteristics of fast, finite time convergence and avoiding singularities \cite{feng2002non}.
Consider the nonlinear sliding surface, $\bm{s} \in \mathbb{R}^6$, as
	\begin{equation} \label{eqSS}
	\bm{s} = \bm{e} + \beta \dot{\bm{e}}^{\alpha},
	\end{equation}
	where $\dot{\bm{e}}^{\alpha}$ is an element-wise exponentiation, i.e., $(\dot{\bm{e}}^{\alpha})_i = (\dot{\bm{e}}_i)^{\alpha}$, $\beta > 0$, $\alpha = \iota_1/\iota_2$, and $\iota_1 > 0$, $\iota_2 > 0$ are adjacent odd numbers such that $1 < \alpha < 2$, and
	\begin{equation} \label{eq:errors}
	\bm{e} = \begin{bmatrix}
	  \tilde{\bm{p}} \\
	\tilde{\bm{\epsilon}}
	\end{bmatrix} \in \mathbb{R}^6, \quad
	\dot{\bm{e}} = \begin{bmatrix}
        \dot{\tilde{\bm{p}}} \\
        \dot{\tilde{\bm{\epsilon}}}
	\end{bmatrix} =
        \begin{bmatrix}
	\dot{\tilde{\bm{p}}} \\
        \frac{1}{2} \left(\tilde{\eta} I + [\tilde{\bm{\epsilon}}] \right) \tilde{\bm{\omega}}
	\end{bmatrix} =
        H
        \begin{bmatrix}
        \dot{\tilde{\bm{p}}} \\
        \tilde{\bm{\omega}}
        \end{bmatrix} \in \mathbb{R}^6, 
	\end{equation}
	where 
    \begin{align*}
        \tilde{\bm{p}} &= \hat{\bm{p}} - \bm{p}_{d}, 
        &\dot{\tilde{\bm{p}}} &= \dot{\hat{\bm{p}}} - \dot{\bm{p}}_{d}, \\
        \tilde{\eta} &= \hat{\eta} \eta_d + \hat{\bm{\epsilon}}^T \bm{\epsilon}_d,
        &\tilde{\bm{\epsilon}} &= -\hat{\eta} \bm{\epsilon}_d + \eta_d \hat{\bm{\epsilon}} - [\hat{\bm{\epsilon}}]\bm{\epsilon}_d, \\
        \tilde{\bm{\omega}} &= \hat{\bm{\omega}} - \bm{\omega}_d,
        &H &= \begin{bmatrix}
        I & 0 \\
        0 & \frac{1}{2} \left(\tilde{\eta} I + [\tilde{\bm{\epsilon}}] \right)
        \end{bmatrix},
    \end{align*}
    and $[\bm{\epsilon}] = \begin{bmatrix}
        0 & -\epsilon_z & \epsilon_y \\
        \epsilon_z & 0 & -\epsilon_x \\
        -\epsilon_y & \epsilon_x & 0
    \end{bmatrix}$ is the skew-symmetric matrix operator. The formulation of the quaternion error, $\tilde{\bm{\epsilon}}$ and $\dot{\tilde{\bm{\epsilon}}}$ can be found in \cite[Ch. 3]{siciliano2008robotics}.

	The main challenge in using a conventional SM controller in real-world applications is the chattering effect. A second-order SM controller is a method of reducing chattering without reducing the controller's performance. The STSM controller is a second-order method that introduces a continuous function of the sliding surface and an integral of the discontinuous function of the sliding surface to minimize chattering. Inspired by the NTSM control designs in \cite{feng2002non} and the STSM control approach in \cite{moreno2012strict}, the proposed controller incorporates quaternion-based orientation error to ensure finite-time convergence in task space and is formulated as follows:
	\begin{align}
	\bm{u} =& \ \bar{\mathcal{M}} \left(\ddot{\bm{x}}_{d} + 
    H^{-1} \left(\frac{-\dot{\bm{e}}^{(2-\alpha)}}{\alpha \beta} - \kappa_1 |S|^{\frac{1}{2}} \textrm{sign} (\bm{s}) + \bm{\nu} \right)
    \right) \nonumber \\
    &+  \bar{\mathcal{C}} + \bar{\mathcal{G}}, \label{eqCTRL} \\
	\dot{\bm{\nu}} =& \ - \kappa_2 \textrm{sign} (\bm{s}), \nonumber
	\end{align}
	where $\dot{\bm{e}}^{(2-\alpha)}$ is an element-wise exponentiation, $|S|^{\frac{1}{2}} = \textrm{diag}(|\bm{s}|^{\frac{1}{2}})$, and
    \begin{equation*}
        \kappa_1 =
        \begin{bmatrix}
            \kappa_{11} & 0 & 0 \\
            0 & \ddots & 0 \\
            0 & 0 & \kappa_{16}
        \end{bmatrix}, \
        \kappa_2 =
        \begin{bmatrix}
            \kappa_{21} & 0 & 0 \\
            0 & \ddots & 0 \\
            0 & 0 & \kappa_{26}
        \end{bmatrix}, \nonumber
    \end{equation*}
    are positive definite matrices. When implementing the controller on the manipulator, it is transformed into joint space, where $\bm{\tau} = J^T \bm{u}$.

    Consider the invertibility of $H$. The determinant of $H$ is given by
        $\textrm{det}(H) = \tilde{\eta} (\tilde{\eta}^2 + \tilde{\bm{\epsilon}}^T\tilde{\bm{\epsilon}})$. 
        The error between two unit quaternions is derived from a quaternion product and must result in a unit quaternion \cite{campa2006kinematic}. According to the unit norm property of quaternions:
        \begin{equation*}
            \tilde{\eta}^2 + \tilde{\bm{\epsilon}}^T \tilde{\bm{\epsilon}} = 1,
        \end{equation*}
        resulting in $\textrm{det}(H) = \tilde{\eta}$.
    
    \begin{assumption}
         It is assumed that the desired orientation trajectory $\bm{\xi}_d$ is sufficiently close to the current robot orientation $\bm{\xi}$ such that $|\tilde{\bm{\epsilon}}| < \bar{\bm{\epsilon}}$, where $\bar{\bm{\epsilon}}$ is a small constant.
     \end{assumption} 
     
         Assumption 1 is ensured by smooth interpolation in \eqref{eq:desquat} and periodic checks of the robot's current pose by the trajectory generator. As a result, the quaternion error $\tilde{\bm{\epsilon}}$ and its derivative $\dot{\tilde{\bm{\epsilon}}}$ can be locally approximated in Euclidean space. This allows for element-wise operations on $\dot{\bm{e}}$ in \eqref{eqSS} and \eqref{eqCTRL}, despite their lack of strict geometric validity on $\mathrm{S}^3$. The proximity between the desired and actual orientations ensures that they do not result in orthogonal conditions or specific alignments that would cause $\textrm{det}(H)$ to be zero. In practice, to ensure numerical robustness, a lower bound on $|\tilde{\eta}|$, such as $|\tilde{\eta}| \geq 0.1$, may be enforced to avoid inverting a poorly conditioned $H$. Therefore, $\textrm{det}(H) = \tilde{\eta} \neq 0$, and $H$ remains invertible.
    
    %
%
\subsection{Adaptive Law}
Adaptive gains offer a dynamic response to uncertainties and disturbances, effectively compensating for them without requiring prior knowledge of their bounds \cite{gonzalez2011variable}. This is important for systems characterized by uncertainties and disturbances that can vary over time. The adaptive law for the controller gains $\kappa_1$ and $\kappa_2$ in \eqref{eqCTRL} is designed as
\begin{align} \label{eq:adptlaw}
	\dot{L}_{i} &= 
		\begin{cases}
			-\eta_{\alpha}, & \text{if } \kappa_{1i} \geq \kappa_{1\text{max}}, \\
			\omega_{\alpha} |s_i| \textrm{sign}(|s_i| - \mu_{\alpha}), & \text{if } \kappa_{1\text{max}} > \kappa_{1i} > \kappa_{1\text{min}}, \\
			\eta_{\alpha}, & \text{if } \kappa_{1i} \leq \kappa_{1\text{min}}, 
		\end{cases} \\
        \kappa_{1i} &= \Omega_{1}\sqrt{\frac{2\gamma L_i}{(1-\theta)\Omega_{2}}}, \qquad \kappa_{2i} = \frac{\theta+1}{1-\theta}L_i,\label{eq:kappa}
\end{align}
%
%
%
where $\bm{L} = \begin{bmatrix}
    L_1 & \cdots & L_6
\end{bmatrix}^T \in \mathbb{R}^6$ is a positive adaptation parameter vector, $\omega_{\alpha}$, $\mu_{\alpha}$, $\eta_{\alpha}$, $\kappa_{1\text{max}}$, $\kappa_{1\text{min}}$, are constant positive adaptive law parameters and $\Omega_1$, $\Omega_2$, $\gamma$, and $\theta$ are constant positive scalar gain parameters that satisfy $0 < \theta < 1$ and
    \begin{equation}
    \Omega_{1} \Gamma_i^2 - 2\frac{\Omega_{2}}{\gamma} \Gamma_i > \frac{1}{4}(\Gamma_i + \Omega_{1} \Gamma_i)^2 - (1+\Omega_{1}) \Omega_{2} \Gamma_i \theta + \Omega_{2}^2,
    \end{equation}
where $\Gamma_i = \alpha \beta \dot{e}_i^{(\alpha - 1)}$. See Remarks 1 and 2 in Section \ref{sec:s_stability} for tuning guidelines.



%
%
\subsection{Non-Singular Terminal Super-Twisting Sliding Mode
Controller Convergence Analysis}\label{sec:s_stability}

This subsection presents the boundedness analysis considering the system \eqref{eqELts} under the controller \eqref{eqCTRL}. First, the main theorem is presented. Then, the closed-loop dynamics are analyzed and reformulated such that the Lyapunov candidate function can be introduced. Next, the convergence of the sliding surface to a bounded region is analyzed. Finally, the tracking bounds of the position and velocity errors are analyzed when the sliding surface is zero and when it converges to a bounded region. 

The main challenge in analyzing the adaptive NT-STSM controller lies in proving boundedness without enforcing fixed gain ratios, such as $\kappa_2 = \frac{1}{4}\Gamma \kappa_1$ used in prior work \cite{sun2024adaptive}. To address this, the nonlinear closed-loop dynamics are transformed into a pseudo-linear form using the coordinate $\bm{\zeta}$, allowing standard Lyapunov tools to be applied \cite{moreno2012strict}. This approach introduces additional tuning parameters $\Omega_1$, $\Omega_2$, $\gamma$, and $\theta$, which define an admissible gain region while preserving convergence, as detailed in Theorem 1. Practical guidelines for selecting these parameters based on the convergence analysis are discussed in Remark 2.

\begin{lemma}
    \cite{bhat2000finite} Assume a positive definite function, $V$, and its derivative satisfies
    \begin{equation}
        \dot{V} \leq -\gamma_1 V^{\gamma_2},
    \end{equation}
    where $\gamma_1 > 0$ and $0 < \gamma_2 < 1$. The function $V$ converges to zero from any initial state within a finite time, determined by
    \begin{equation}
        t_f(x_0) \leq \frac{V^{1-\gamma_2}(x_0)}{\gamma_1 (1 - \gamma_2)}.
    \end{equation}
\end{lemma}
    \begin{assumption} \label{ass:err}
        The lumped disturbances on the system, $\bm{\Pi} = H \bar{\mathcal{M}}^{-1} \bm{f}_d = H J \mathcal{M}^{-1} \bm{\tau}_d$, are first-order differentiable and there exists a vector of positive constants, $\bm{\delta} \in \mathbb{R}^{6}$, such that $|\dot{\bm{\Pi}}| \leq \bm{\delta}$ is globally bounded. 
    \end{assumption}

    Since $H$ appears in $\bm{\Pi}$, Assumption 2 implicitly assumes the rate of change of $H$ is bounded, ensuring that the orientation error does not change too rapidly due to disturbances or uncertainties.

    \begin{assumption}
        The measurement noise in $\bm{x}$ and $\dot{\hat{\bm{x}}}$ is bounded by a constant $\epsilon_c$.
    \end{assumption}
\begin{assumption}
    The initial state $(\bm{x}(0), \dot{\hat{\bm{x}}}(0))$ lies within a region where the control law \eqref{eqCTRL} is well-defined and the adaptation dynamics are initialized such that $\kappa_{1\min} \leq \kappa_1(0) \leq \kappa_{1\max}$, where $\kappa_{1\min}$ and $\kappa_{1\max}$ are known constants.
\end{assumption}
\begin{theorem}\label{theorem:1}
    Consider the system dynamics \eqref{eqELts} satisfying Assumption \ref{ass:err} with the control input \eqref{eqCTRL}. If the variable gains, $\kappa_1$ and $\kappa_2$ are selected as in \eqref{eq:kappa}, then in the presence of Lebesgue-measurable noise of maximal magnitude $\epsilon_c$, while selecting $\mu_{\alpha} > \epsilon_c$, a vicinity of the origin defined by $\bm{s} \leq \mu_{\alpha}$ is globally and robustly finite time stable for every value of perturbation derivative $|\dot{\bm{\Pi}}| \leq \bm{\delta}$. A trajectory starting at $\bm{s}(0)$ will converge to the region $|\bm{s}| \leq \mu_{\alpha}$ from any initial condition $|\bm{s}(0)| > \mu_{\alpha}$ in finite time. The upper bound on the time to reach this domain can be estimated as
    \begin{equation}
       t_f(\bm{s}(0)) \leq \frac{2}{\vartheta}V_1^{\frac{1}{2}}(\bm{s}(0)),
    \end{equation}
    where 
    \begin{align} \label{eq:expandV1}
        V_1 =& |\bm{s}| - 2 \sqrt{\frac{(1-\theta) \Omega_2}{2 \gamma L}} |\bm{s}|^{\frac{1}{2}} \textrm{sign}(\bm{s}) (\bm{\nu} + \bm{\Pi}) \nonumber \\
        &+ \frac{(1-\theta)\Omega_2}{2 L}(\bm{\nu} + \bm{\Pi})^2 + \frac{1}{2} (L-L^*)^2,
    \end{align}
    and $\vartheta = \textrm{min}\{ \chi, \sqrt{2} \omega_{\alpha} \mu_{\alpha} \}$, $\chi = \frac{\lambda^{\frac{1}{2}}_{\textrm{min}}(P)\lambda_{\textrm{min}}(\tilde{Q}_R)}{\lambda_{\textrm{max}}(P)}$, and $L^* \geq L$ is a positive constant. Lastly, the finite time tracking control convergence is achieved in a finite time according to 
    %
    %
    \begin{equation}
        t_f(\bm{e}) \leq \frac{2\alpha}{\beta^{-\frac{1}{\alpha}}(\alpha-1)}\frac{V_3(\bm{e}(0))^{\frac{\alpha-1}{2\alpha}}}{2^{\frac{1+\alpha}{2\alpha}}},
    \end{equation}
    where $V_3 = \frac{1}{2}\bm{e}^2$.
    In the presence of measurement noise, the tracking errors converge to the following regions
    \begin{equation}\label{eq:regions}
    |{\bm{e}}| \leq 2\mu_{\alpha}, \ \ |{\dot{\bm{e}}}| \leq \left(\frac{\mu_{\alpha}}{\beta}\right)^{\frac{1}{\alpha}}.
\end{equation}
    \end{theorem}

\textbf{Proof:} 

\subsubsection{Closed-loop Dynamics}
This section introduces the closed-loop dynamics of the manipulator and arranges the system into an appealing form.

Given the system \eqref{eqELts} with the controller \eqref{eqCTRL}, the closed-loop dynamics can be written as
	%
    \begin{equation} \label{eq:CLdyn}
	\ddot{\bm{x}} = \ddot{\bm{x}}_{d} + 
    H^{-1} \left(\frac{-\dot{\bm{e}}^{(2-\alpha)}}{\alpha \beta} - \kappa_1 |S|^{\frac{1}{2}} \textrm{sign} (\bm{s}) + \bm{\nu} \right)
    + \bar{\mathcal{M}}^{-1} \bm{f}_{d}.
    \end{equation}
    The acceleration error is expressed as in \cite{campa2006kinematic}
    \begin{equation}
        \ddot{\bm{e}} = \begin{bmatrix}
        \ddot{\tilde{\bm{p}}} \\
        \ddot{\tilde{\bm{\epsilon}}}
        \end{bmatrix} =
        \begin{bmatrix}
        \ddot{\tilde{\bm{p}}} \\
        \left(\tilde{\eta} I + [\tilde{\bm{\epsilon}}] \right) \dot{\tilde{\bm{\omega}}}
        \end{bmatrix},
    \end{equation}
    where $\dot{\tilde{\bm{\omega}}} = \dot{\bm{\omega}} - \dot{\bm{\omega}}_d$.
    Rearranging \eqref{eq:CLdyn} results in
    \begin{align}
        \ddot{\bm{e}} &= H (\ddot{\bm{x}} - \ddot{\bm{x}}_{d}) \nonumber \\
        &= \frac{-\dot{\bm{e}}^{(2-\alpha)}}{\alpha \beta} - \kappa_1 |S|^{\frac{1}{2}} \textrm{sign} (\bm{s}) + \bm{\nu} + H \bar{\mathcal{M}}^{-1} \bm{f}_d. \label{eq:eddot}
    \end{align}
    
    %

    The derivative of the sliding surface \eqref{eqSS} is derived as
    \begin{align}
    \dot{\bm{s}} &= \dot{\bm{e}} + \alpha \beta  \textrm{diag}(\dot{\bm{e}}^{(\alpha - 1)}) \ddot{\bm{e}} \nonumber \\
    &= \dot{\bm{e}} + \bm{\Gamma} \left(\frac{-\dot{\bm{e}}^{(2-\alpha)}}{\alpha \beta} - \kappa_1 |S|^{\frac{1}{2}} \textrm{sign} (\bm{s}) + \bm{\nu} + H \bar{\mathcal{M}}^{-1} \bm{f}_d \right) \nonumber \\
    &= \bm{\Gamma} (- \kappa_1 |S|^{\frac{1}{2}} \textrm{sign} (\bm{s}) + \bm{\nu} + \bm{\Pi}).
    \end{align}
    Due to the constraints on $\alpha$ and $\beta$ in \eqref{eqSS}, $\bm{\Gamma} = \alpha \beta \textrm{diag}(\dot{\bm{e}}^{(\alpha - 1)}) \geq 0$. Introducing new states as $\bm{y}_1 = \bm{s}$ and $\bm{y}_2 = \bm{\nu} + \bm{\Pi}$,
    %
    %
    we have
    \begin{align}
        \dot{\bm{y}}_1 &= \dot{\bm{s}} = \Gamma (- \kappa_1 |Y_1|^{\frac{1}{2}} \textrm{sign} (\bm{y}_1) + \bm{y}_2),\label{eq:STA_a} \\
        \dot{\bm{y}}_2 &= \dot{\bm{\nu}} + \dot{\bm{\Pi}} = - \kappa_2 \textrm{sign} (\bm{y}_1) + \dot{\bm{\Pi}}, \label{eq:STA_b}
    \end{align}
    where $|Y_1|^{\frac{1}{2}} = \textrm{diag}(|\bm{y}_1|^{\frac{1}{2}}) \in \mathbb{R}^{6 \times 6}$.

    Define $\bm{\zeta}$ as in \cite{moreno2008lyapunov}, taking the $i$-th Cartesian dimension,
    \begin{equation} \label{eq:zetadyn}
        \bm{\zeta} = \begin{bmatrix}
        \zeta_{1} & \zeta_{2}
    \end{bmatrix}^T = \begin{bmatrix}
        |y_{1i}|^{\frac{1}{2}} \textrm{sign} (y_{1i}), & y_{2i}
    \end{bmatrix}^T.
    \end{equation}
    For clarity, assume the $i$-th element for each term in the remainder of this section. Denoting $y_1 \triangleq y_{1i}, \ y_2 \triangleq y_{2i}, \ \kappa_1 \triangleq \kappa_{1i}, \ \kappa_2 \triangleq \kappa_{2i}, \ \Gamma \triangleq \Gamma_i,  \ \dot{\Pi} \triangleq \Pi_i$, then
    \begin{align}
        \dot{\zeta}_{1} &= \frac{1}{2} |y_{1}|^{-\frac{1}{2}} \dot{y}_{1} = \frac{1}{2} |y_{1}|^{-\frac{1}{2}} \Gamma (- \kappa_{1} |y_{1}|^{\frac{1}{2}} \textrm{sign} (y_{1}) + y_{2}) \nonumber \\
        &= 
        |y_{1}|^{-\frac{1}{2}} \frac{1}{2} \Gamma (-\kappa_{1}  \zeta_{1} + \zeta_{2}), \label{eq:y1} \\
        \dot{\zeta}_2 &= -\kappa_2 \textrm{sign}(y_{1}) + \dot{\Pi} = |y_{1}|^{-\frac{1}{2}} (-\kappa_2 |y_{1}|^{\frac{1}{2}} \textrm{sign}(y_{1})) + \dot{\Pi} \nonumber \\
        &= |y_{1}|^{-\frac{1}{2}} (-\kappa_2 \zeta_{1}) + \dot{\Pi}. \label{eq:y2}
    \end{align}
    Introducing a new variable $\pi(t,y) = \dot{\Pi}\textrm{sign}(y_{1})$, \eqref{eq:y1} and \eqref{eq:y2} can be arranged into
    \begin{equation} \label{eq:zetadot}
        \dot{\bm{\zeta}} = |y_{1}|^{-\frac{1}{2}} A \begin{bmatrix}
            \zeta_{1} \\ \zeta_{2}
        \end{bmatrix},
    \end{equation}
    where $A = \begin{bmatrix}
            -\frac{1}{2} \kappa_{1} \Gamma & \frac{1}{2}\Gamma \\
            -(\kappa_{2}-\pi) & 0 
        \end{bmatrix}$, similar to the form in \cite{moreno2012strict}.
        
    %
    
    \subsubsection{Sliding Mode Dynamics Boundedness Analysis}
    The following analysis considers the boundedness of the sliding mode dynamics \eqref{eq:zetadot} while considering Lebesgue-measurable noise and adaptive gain selection \eqref{eq:adptlaw}.
    
    Consider the following Lyapunov function, noting that it is the condensed form of \eqref{eq:expandV1}:
    \begin{align}
        V_1 &= V_2 + \frac{1}{2} (L-L^*)^2, \label{eq:lya}
    \end{align}
where $L^*$ is the upper bound of the gain values $L\leq L^*$, and
\begin{equation}
V_2 =  \bm{\zeta}^T P \bm{\zeta}.
\end{equation}
    %
    The matrix $P$ is symmetric and positive definite as 
    \begin{equation}\label{eq:P_matrix}
    P = \begin{bmatrix}
        p_{11} & p_{12} \\
        p_{21} & p_{22}
    \end{bmatrix}.
\end{equation}

    %
    %
    %
    As in \cite{moreno2008lyapunov}, since $V_2$ is not locally Lipschitz, conventional versions of the Lyapunov theorem cannot be used here. However, $V_2$ is continuous along the state trajectories of $\bm{y_1} = \bm{s}$ and therefore the boundedness can be shown using Zubov's Theorem \cite[Theorem 20.2]{poznyak2009advanced}. Following Rayleigh's Theorem as in \cite{sun2024adaptive}, $V_2$ is positive definite and radially unbounded:
    \begin{equation}\label{eq:P_condition}
        \lambda_{\textrm{min}}(P) ||\bm{\zeta}||_2^2 \leq V_2 \leq \lambda_{\textrm{max}}(P) ||\bm{\zeta}||_2^2,
    \end{equation}
    where $||\bm{\zeta}||_2^2 = |y_{1}| + y^2_2$.
    Differentiating $V_2$ along the trajectories of the closed-loop dynamics \eqref{eq:zetadot}
    \begin{equation}
        \dot{V}_2 = -|y_1|^{-\frac{1}{2}} \bm{\zeta}^T \tilde{Q} \bm{\zeta}. \label{eq:V1dot}
    \end{equation}
    %
Let $\tilde{Q}$ be defined as
        \begin{equation} \label{eq:Q}
            -\Tilde{Q} = A^T P + PA, \ \ \Tilde{Q}=
           \begin{bmatrix}
            \Tilde{q}_{11} & \Tilde{q}_{12} \\
            \Tilde{q}_{21} & \Tilde{q}_{22}
        \end{bmatrix},
        \end{equation}
which is computed as
\begin{equation} 
{\tilde{Q}}:
\begin{cases}
    \tilde{q}_{11} = \kappa_1 \Gamma p_{11} + 2p_{12}(\kappa_2 - \pi), \\
    \tilde{q}_{12} = \tilde{q}_{21} = \frac{1}{2}\kappa_1 \Gamma p_{12} + p_{22}(\kappa_2 - \pi) - \frac{1}{2}\Gamma p_{11}, \\
    \tilde{q}_{22} = -\Gamma p_{12}.
\end{cases}
\end{equation}

Although \eqref{eq:Q} resembles a Linear Time Invariant (LTI) Lyapunov equation, it results from a coordinate transformation of the nonlinear dynamics, allowing classical stability analysis to be applied \cite{moreno2012strict}. Given that $\pi$ is bounded, $V_2$ is positive definite and $\dot{V}_2$ is negative definite for every value of the perturbation derivative $|\dot{\bm{\Pi}}| \leq \bm{\delta}$ subject to
\begin{align}\label{eq:gain_cond}
    \begin{split}
        &{p}_{11} = 1, \ \ \ {p}_{22} > {p_{12}}^2, \ \ \ 
        {p}_{12} < 0, \ \ \
        \Gamma > 0, \ \ \ 
        \forall t \geq 0,
    \end{split} \\
    &\kappa_1\Gamma^2p_{11}p_{12} + \frac{1}{4}(\Gamma p_{11} - \kappa_1\Gamma p_{12})^2 + 2\Gamma p_{12}^2(\kappa_2 - \pi) \nonumber \\
    &- (p_{11} - \kappa_1p_{12})\Gamma p_{22}(\kappa_2 - \pi) + (\kappa_2 - \pi)^2p_{22}^2 < 0. \label{eq:condition}
\end{align}
%

%
%

\begin{figure}
    \centering
    \includegraphics[width=\linewidth]{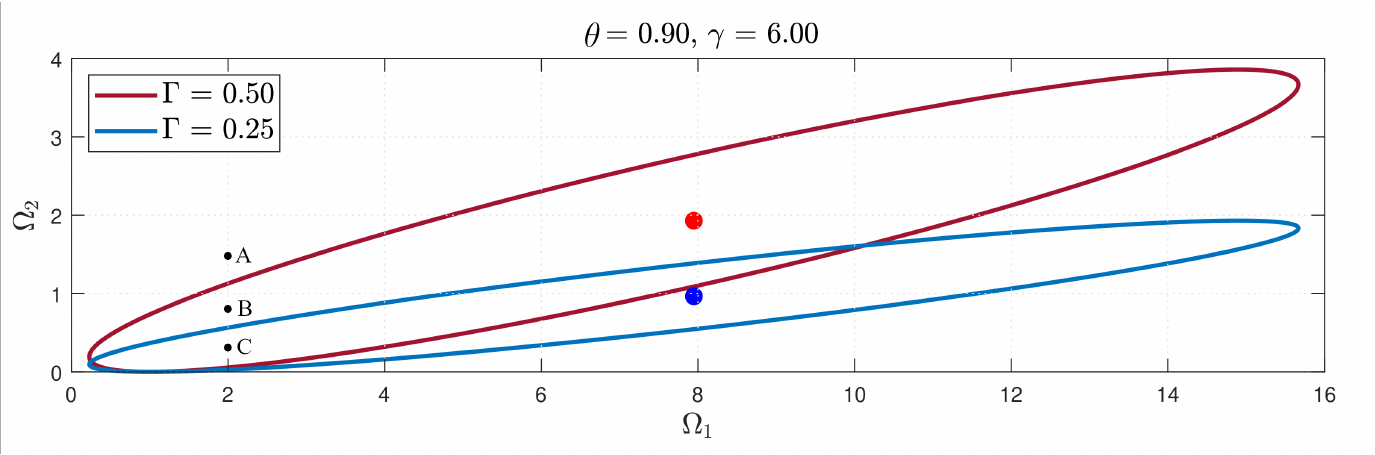}
    \caption{Ellipses expressing the stable boundary of \eqref{eq:condition_final2} for different values of $\Gamma$. $A$ is not bounded for the given parameters, $B$ is bounded for $\Gamma =0.5$ and unbounded for $\Gamma =0.25$, and $C$ is bounded for both $\Gamma =0.5$ and $\Gamma =0.25$.}
    \label{fig:ellipse}
\end{figure}

Noting that $(\kappa_2 - L) \leq (\kappa_2 - \pi) \leq (\kappa_2 + L)$, with $L$ defined in \eqref{eq:adptlaw}. When $L < \pi$, then $|s_i| < \mu_{\alpha}$ and from \eqref{eq:adptlaw}, $L$ will increase until the sliding mode occurs or $L > \pi$ is established. To maintain the sliding mode and ensure the boundedness around the origin for perturbation $\pi(t,y) = L$, $\kappa_2 > L$ must be satisfied. Therefore, the condition \eqref{eq:condition} will be satisfied if
\begin{multline}\label{eq:condition_satisfied}
    \kappa_1\Gamma^2p_{12} + \frac{1}{4}(\Gamma - \kappa_1\Gamma p_{12})^2 + 2\Gamma p_{12}^2(\kappa_2 + L) \\- (1 - \kappa_1p_{12})\Gamma p_{22}(\kappa_2 - L) + (\kappa_2 + L)^2p_{22}^2 < 0.
\end{multline}
Define
\begin{equation}\label{eq:ellipse_1st}
    \begin{cases}
    \gamma = \frac{p_{22}}{p_{12}^2}, \ \ \theta = \frac{\kappa_2 - L}{\kappa_2 + L},\\ \Omega_1 = -\kappa_1 p_{12}, \ \
    \Omega_2 = p_{22}(\kappa_2 +L),  
    \end{cases}
\end{equation}
and the condition \eqref{eq:condition_satisfied} is satisfied if
\begin{multline}\label{eq:condition_final2}
    \Omega_1 \Gamma^2 - 2\frac{\Omega_2}{\gamma}\Gamma > \frac{1}{4}(\Gamma + \Omega_1 \Gamma)^2 - (1+\Omega_1) \Omega_2\Gamma\theta + \Omega_2^2,\\ \ \ \ 0 < \theta < 1,
\end{multline}
is met. The first inequality in \eqref{eq:condition_final2} represents the interior of a size-variable ellipse in the plane of $\Omega_1$ and $\Omega_2$, of which the size is determined by $\gamma$, $\theta$, and $\Gamma$. Functions $\Omega_1,\Omega_2 \in\mathbb{R}^+$ must be selected to satisfy \eqref{eq:condition_final2}, which requires $\gamma > 1$, and is always possible if $\gamma\theta>1$. Fig.~\ref{fig:ellipse} illustrates this elliptical stability region and how it evolves with $\Gamma$.

Given the parameters $\theta, \gamma, \Omega_1, \Omega_2$ that satisfy \eqref{eq:condition_final2}, the components of \eqref{eq:P_matrix} that satisfy \eqref{eq:gain_cond} and \eqref{eq:condition} are given as
\begin{equation}\label{eq:P_components}
{P}:
\begin{cases}
        {p}_{11} = 1, \\
        {p}_{12} = p_{21} = -\sqrt{\frac{(1-\theta)\Omega_2}{2\gamma L}} = -\sqrt{\frac{p_{22}}{\gamma}}, \\
        {p}_{22} = \frac{(1-\theta)\Omega_2}{2L}.
\end{cases}
\end{equation}
The variable gains $\kappa_1$ and $\kappa_2$, provided values of $\theta$ and $\gamma$, are given as
\begin{equation}\label{eq:variable_kappas}
        \kappa_1 = \Omega_{1}\sqrt{\frac{2\gamma L}{(1-\theta)\Omega_{2}}}, \ \ \kappa_2 = \frac{\theta+1}{1-\theta}L,
\end{equation}
with $\Omega_1$ and $\Omega_2$ selected as any points within the ellipse \eqref{eq:condition_satisfied}, which exist in the first quadrant. This selection assures robust, finite-time boundedness about the origin of the super-twisting algorithm.

\textit{Case 1a:}\label{Case 1} For the case when $\Gamma \neq 0$, implying $\dot{\bm{e}} \neq 0$, using \eqref{eq:V1dot} we have that
\begin{equation}\label{eq:STA_lyapunov_comp}
        \dot{V}_2 \leq -|y_1|^{-\frac{1}{2}}\lambda_{min}(\tilde{Q}_R)||\boldsymbol{\zeta}||_2^2,
\end{equation}
where $\tilde{Q}_R \in \mathbb{R}^+$ is computed as
\begin{equation}\label{eq:Q_R}
{\tilde{Q}_R}:
\begin{cases}
        \tilde{q}_{R_{11}} =   \kappa_1 \Gamma - 2\Omega_{2}\frac{p_{12}}{p_{22}} + (1+\Omega_{1})\Omega_{2}\frac{1-\theta}{p_{12}}, \\
        \tilde{q}_{R_{12}} = \tilde{q}_{R_{21}} = \Omega_{2} - \frac{1}{2}(\Gamma + \Gamma \Omega_{1}), \\
        \tilde{q}_{R_{22}} = -\Gamma p_{12},
\end{cases}
\end{equation}
and remains positive definite as long as \eqref{eq:condition_final2} holds.


By arranging \eqref{eq:P_condition} such that $-||\bm{\zeta}||^2_2 \leq -\frac{V_2}{\lambda_{\textrm{max}}(P)}$, Eq. \eqref{eq:STA_lyapunov_comp} becomes
\begin{equation*}
    \dot{V}_2 \leq -|y_1|^{-\frac{1}{2}} \frac{\lambda_{\textrm{min}}(\tilde{Q}_R)}{\lambda_{\textrm{max}}(P)} V_2.
\end{equation*}
Using the fact that $|y_1|^{\frac{1}{2}} \leq ||\bm{\zeta}||_2$, it follows that $-|y_1|^{-\frac{1}{2}} \leq -(||\bm{\zeta}||_2)^{-1}$ and
\begin{equation*}
    \dot{V}_2 \leq -(||\bm{\zeta}||_2)^{-1} \frac{\lambda_{\textrm{min}}(\tilde{Q}_R)}{\lambda_{\textrm{max}}(P)} V_2.
\end{equation*}
Rearranging \eqref{eq:P_condition} such that $-\frac{1}{||\bm{\zeta}||_2} \leq -\frac{\lambda^{\frac{1}{2}}_{\textrm{min}}(P)}{V_2^{\frac{1}{2}}}$, we have
\begin{equation}
    \dot{V}_2 \leq - \frac{\lambda^{\frac{1}{2}}_{\textrm{min}}(P)\lambda_{\textrm{min}}(\tilde{Q}_R)}{\lambda_{\textrm{max}}(P)} V^{\frac{1}{2}}_2 \leq - \chi V^{\frac{1}{2}}_2, \label{eq:manuel_final}
\end{equation}
where $\chi = \frac{\lambda^{\frac{1}{2}}_{\textrm{min}}(P)\lambda_{\textrm{min}}(\tilde{Q}_R)}{\lambda_{\textrm{max}}(P)} > 0$. From \cite[Lemma 2]{Lemma_ref}, it is possible to assume that there exists a positive constant $L^*$ such that $L\leq L^*$ and $\forall t > 0$. Consider the Lyapunov function \eqref{eq:lya}, taking the derivative along the trajectories of the perturbed system while using \eqref{eq:manuel_final} and the gain $L_i$ in \eqref{eq:adptlaw} gives
\begin{align}
        \dot{V}_1 &= \dot{V}_2 + \dot{L}(L-L^*)  \\
        &\leq - \chi V^{\frac{1}{2}}_2 - \underbrace{\omega_{\alpha} |s| \textrm{sign}(|s| - \mu_{\alpha})|L-L^*|}_{\substack{\Upsilon}}, \label{eq:vdot}
    \end{align}
Suppose that $|s| > \mu_{\alpha}$, this gives
    \begin{equation}
        \dot{V}_1 \leq - \chi V^{\frac{1}{2}}_2 -  \omega_{\alpha} \mu_{\alpha}|L-L^*|.
    \end{equation}
Note that $\omega_{\alpha} > 0$ and $\mu_{\alpha} > 0$. Jensen's inequality for two variables, with $a \geq 0$, $b \geq 0$, and $0 < m < n$, implies that $(a^n + b^n)^{\frac{1}{n}} \leq (a^m + b^m)^{\frac{1}{m}}$. Setting $n = 2$, where $m = 1$
    \begin{equation}
        \dot{V}_1 \leq - (\chi^2 V_2 + 2 \omega_{\alpha}^2 \mu_{\alpha}^2(\frac{1}{2}|L-L^*|^2))^{\frac{1}{2}} \leq - \vartheta V_1^{\frac{1}{2}},
    \end{equation}
where $\vartheta = \textrm{min}\{ \chi, \sqrt{2} \omega_{\alpha} \mu_{\alpha} \}$. Therefore, the finite time converge to the region $|s| \leq \mu_{\alpha}$ is guaranteed from any initial condition $|s(0)| > \mu_{\alpha}$. In this case, the states $\dot{e}$ will be driven onto the sliding manifold $s$ before the states $\dot{e}$ converge to the origin within a domain of $|s| \leq \mu_{\alpha}$. Using Lemma 1, the upper bound on the reaching time is given as 

\begin{equation}\label{eq:s_reachingTime}
    t_f(y_0) \leq \frac{2}{\vartheta}V_1^{\frac{1}{2}}(y_0),
\end{equation}
where $y_0$ is the initial value of $y$. Experimental evaluation has shown that a value of $\mu_{\alpha}$ can be selected slightly higher than the measurement noise level on the corresponding position or rotation measurement output ($\epsilon_c<\mu_{\alpha}$) to attain a dynamic reduction of controller gains.

Suppose that $|s| < \mu_{\alpha}$. In this scenario, the function $\Upsilon$ in \eqref{eq:vdot} can be negative. Consequently, $\Dot{V}_1$ would be sign-indefinite, and the closed-loop boundedness of the $s$-dynamics cannot be concluded. However, when $s$ escapes the boundary ($s> \mu_{\alpha}$), under \eqref{eq:adptlaw}, $L$ immediately begins to increase again to keep $s$ in the domain. Thus, $L$ continues changing its value, and $s$ remains around the boundary layer limit $|s| = \mu_{\alpha}$. In conclusion, $s$ initially converges to the domain $|s| < \mu_{\alpha}$ in finite time and subsequently remains in the region $|s| \leq \mu_{\alpha}$, verifying the sliding mode boundedness \cite{smith2024adaptive}.

\textit{Case 1b:}\label{Case 2}
Consider the case when $\dot{e} = 0$ and $s \neq 0$. From \eqref{eq:condition_final2} and \eqref{eq:ellipse_center}, as $\Gamma \rightarrow 0$ this vertically compresses the ellipse represented by \eqref{eq:condition_final2} towards the $\Omega_1$ axis, shown in Fig.~\ref{fig:ellipse}. This may result in an obstruction of the reachability of \eqref{eqSS}.
%



%
For the case when $\dot{e} = 0$ and $s \neq 0$, \eqref{eq:eddot} becomes
    \begin{align}
        \ddot{e} &= - \kappa_1 |s|^{\frac{1}{2}} \textrm{sign} (s) + \nu + \Pi \nonumber \\
        &= - \kappa_1 |s|^{\frac{1}{2}} \textrm{sign} (s) - \int_0^t (\kappa_2 \textrm{sign} (s) - \dot{\Pi}) dt.
    \end{align}
    Since $\kappa_2  > |\dot{\Pi}|$, ensuring boundedness around the origin, $\ddot{e} \neq 0$ indicates that $\dot{e} = 0$ is not an attractor in the reaching phase. Hence, the reachability of the sliding manifold \eqref{eqSS} to a boundary in the finite time as $t_f(y_0)$ in \eqref{eq:s_reachingTime} is still guaranteed. Therefore it will not interfere with the convergence of $V_1$.


\subsubsection{State Convergence Analysis}
The following analyses consider the state tracking stability and convergence of $e$ and $\dot{e}$, first without noise and second with noise, when the absolute second order sliding mode (SOSM), $\dot{s}=s=0$, is not possible.


\textit{Case 2a:}
Consider the case with no measurement noise. From \eqref{eqSS}, we have
\begin{equation}\label{eq:manifold_arranged}
    e =  s - \beta |\dot{e}|^{\alpha} \textrm{sign}(\dot{e}).
\end{equation}
When SOSM ($\dot{s}=s=0$) occurs in finite time in the absence of measurement noise, \eqref{eq:manifold_arranged} reduces to 
\begin{equation}\label{eq:TSM_form}
    e = - \beta |\dot{e}|^{\alpha} \textrm{sign}(\dot{e}).
\end{equation}
Considering the definition of $\alpha \in (1,2)$: $\textrm{diag}(|\dot{e}|^{\alpha}) \textrm{sign}(\dot{e}) = \dot{e}^{\alpha}$  for $\dot{e} > 0$ and $\dot{e} < 0$, the closed-loop dynamics is
\begin{equation}\label{eq:final_cl_sys}
    \dot{e} = -\beta^{-\frac{1}{\alpha}}(e)^{\frac{1}{\alpha}},
\end{equation}
and the finite time convergence of $e$ and $\dot{e}$ is achieved.

Consider the Lyapunov function candidate as
\begin{equation} \label{eq:final_lyapnov_obserror}
    V_3 = \frac{1}{2}e^2.
\end{equation}
Taking the time derivative of \eqref{eq:final_lyapnov_obserror} and substituting \eqref{eq:final_cl_sys} gives
\begin{align}
        \dot{V}_3  &= {e} \dot{e}  = {e}\left(-\beta^{-\frac{1}{\alpha}}(e)^{\frac{1}{\alpha}}\right) =-\beta^{-\frac{1}{\alpha}}(e)^{\frac{1+\alpha}{\alpha}} \nonumber \\
        &\leq -\beta^{-\frac{1}{\alpha}} 2^{\frac{1+\alpha}{2\alpha}}V_3^{\frac{1+\alpha}{2\alpha}}.
\end{align}
Therefore, we obtain that $e$ and $\dot{e}$ can converge to the origin along the sliding 
surface in finite time as long as $\beta > 0$ and $1<\alpha<2$. With $\alpha \in (1,2)$, this implies $\frac{\alpha + 1}{2\alpha}\in (\frac{3}{4}, 1)$. Using Lemma 1, the upper bound on the convergence time is
\begin{equation}
    t_f(e) \leq \frac{2\alpha}{\beta^{-\frac{1}{\alpha}}(\alpha-1)}\frac{V_3(e(0))^{\frac{\alpha-1}{2\alpha}}}{2^{\frac{1+\alpha}{2\alpha}}}.
\end{equation}

\textit{Case 2b:}
In the presence of bounded Lebesgue-measurable noise, it was shown in Section \ref{sec:s_stability} that the sliding variable $s$ will converge to a region $|s| \leq \mu_{\alpha}$, which is guaranteed from any initial condition $|s(0)| > \mu_{\alpha}$.  With these considerations, from \eqref{eqSS} we get
\begin{equation}\label{eq:final_cl_sys_obserror}
    {e} + \beta|\dot{e}|^{\alpha}\textrm{sign}(\dot{e})= s, \ \ |s| \leq \mu_{\alpha},
\end{equation}
which can be rewritten, when $|\dot{e}| \neq 0$, as
\begin{equation}\label{eq:adjusted_TSM}
    {e} + \underbrace{\left(\beta - \frac{s}{|\dot{e}|^{\alpha}\textrm{sign}(\dot{e})}\right)}_{\substack{\pi}}|\dot{e}|^{\alpha}\textrm{sign}(\dot{e})= 0.
\end{equation}
When $\pi > 0$, then Eq. \eqref{eq:adjusted_TSM} keeps the NTSM form considered in \eqref{eq:TSM_form} and the finite time convergence is maintained and the error is bounded. Consequently, from $\pi$, the velocity tracking error converges to a region 
\begin{equation}\label{eq:vel_region}
    |\dot{e}| \leq \left(\frac{\mu_{\alpha}}{\beta}\right)^{\frac{1}{\alpha}}.
\end{equation}

Additionally, from \eqref{eq:final_cl_sys_obserror}$-$\eqref{eq:vel_region}, we can conclude that the position error will converge to a region
\begin{equation}\label{eq:position_region}
    |{e}| \leq \beta|\dot{e}|^{\alpha} + |s| \leq 2\mu_{\alpha},
\end{equation}
in finite time. This concludes the proof.

\begin{remark}
In practical applications, $\mu_\alpha$ should be selected based on the desired steady-state tracking error and the level of measurement noise. Specifically, it should satisfy $\mu_\alpha > \epsilon_c$, where $\epsilon_c$ is the magnitude of the maximum expected noise. However, as $\epsilon_c$ is often difficult to determine, $\mu_\alpha$ can be iteratively decreased until an acceptable accuracy is reached or the adaptive gains continuously increase due to noise. Typical values for $\mu_\alpha$ may range from $0.001$ to $0.01$ depending on hardware and application requirements. The adaptation rate $\omega_\alpha$ provides a trade-off where larger values increase responsiveness to dynamic disturbances and noise but may induce undesirable oscillations. The gain limits $\kappa_{1\text{max}}$ and $\kappa_{1\text{min}}$ constrain the adaptive gain range. $\kappa_{1\text{min}}$ should be set high enough to reject minor disturbances, while $\kappa_{1\text{max}}$ should be selected based on actuator torque limits to avoid saturation. The parameter $\eta_{\alpha}$ can be set to a small value to contain the adaptive gains within the limits.
\end{remark}

\begin{remark}
The boundedness proof for the adaptive NT-STSM controller offers more flexibility in tuning compared to recent similar applications \cite{sun2024adaptive}. The proposed method allows for a range of acceptable ratios between $\kappa_1$ and $\kappa_2$ within defined regions $\Omega_1$ and $\Omega_2$, in contrast to the strict relationship $\kappa_2 = \frac{1}{4} \Gamma \kappa_1$ used in \cite{sun2024adaptive}. The relationship between $\kappa_1$ and $\kappa_2$ imposed by Theorem \ref{theorem:1} is obtained by rearranging \eqref{eq:variable_kappas}:
\begin{equation}
    \kappa_2 = \frac{\Omega_2 (\theta + 1)}{2 \gamma \Omega_1^2} \kappa_1^2,
\end{equation}
offering additional freedom through parameters $\Omega_1$, $\Omega_2$, $\gamma$, $\theta$.
 
 Conservatively, $\Omega_1$ and $\Omega_2$ can be selected as the center of the ellipse represented by 
\eqref{eq:condition_final2}, which is given as
\begin{equation}\label{eq:ellipse_center}
    \Omega_{1c} = \frac{\gamma - 2\theta + \gamma\theta^2}{\gamma(1-\theta^2)}, \ \ \Omega_{2c} = \Gamma \frac{\theta \gamma - 1}{\gamma(1-\theta^2)}.
\end{equation}
However, as illustrated in Fig.~\ref{fig:ellipse} and \eqref{eq:ellipse_center}, as $\bm{\Gamma}$ decreases, the ellipse is compressed and the viable region for $\Omega_1$ and $\Omega_2$ shrinks. From this observation, choosing small values of $\Omega_1$ and $\Omega_2$, located in the bottom left region of the ellipse, ensures stability as $\bm{\Gamma}$ decreases. While this is more restrictive than the traditional STSM control \cite{moreno2012strict}, in which any values of $\Omega_1$ and $\Omega_2$ within the ellipse can be chosen, it is less restrictive than the relationship in \cite{sun2024adaptive} and allows for the practical implementation of NT-STSM control on hardware. Additionally, it is recommended to choose larger values of $\gamma$ and $\theta$ to maintain a large ellipse boundary as $\bm{\Gamma}$ decreases.

Moreover, although the stability condition \eqref{eq:condition_final2} becomes more difficult to satisfy as $\Gamma \rightarrow 0$, the proposed controller remains robust in practice. This is because the proposed NT-STSM controller drives the velocity error to a bounded region \eqref{eq:vel_region} rather than to zero. Since $\Gamma = \alpha \beta |\dot{\bm{e}}|^{\alpha-1}$, it is also driven to a bounded region $\Gamma \in \left( \alpha \mu_{\alpha}^{\frac{\alpha-1}{\alpha}}, \Gamma_{\max} \right]$,
where $\Gamma_{\max}$ is based on the maximum allowable velocity of the system. Additionally, due to measurement noise and persistent disturbances, $\dot{\bm{e}}$ does not vanish in practical implementations. The adaptive law 
\eqref{eq:adptlaw} is active only outside a boundary layer defined by $\mu_\alpha$, enforcing an effective operating region for $\Gamma$ that is lower bounded by $\Gamma_{\min}=\alpha \mu_{\alpha}^{\frac{\alpha-1}{\alpha}}$. This ensures that adaptation stops near the sliding surface and the stable region defined by the inequality \eqref{eq:condition_final2} remains feasible.
\end{remark}


\begin{remark}
Although the boundedness proof is conducted in continuous time, the controller is implemented digitally. This approach is consistent with standard practice in nonlinear control, where sufficiently high sampling rates ensure that discrete-time implementations closely follow continuous-time behavior \cite{slotine1991nonlinear, laila2002open}. Following the results in \cite{laila2002open}, if the continuous-time closed loop satisfies a Lyapunov inequality, then there exists a computable, conservative maximum allowable sampling period $\Delta t^\star$ such that for all $0<\Delta t<\Delta t^\star$ the sampled-data implementation preserves the same inequality. Additionally, the asymptotic tracking error of discrete-time STSM control scales as $\mathcal{O}(\Delta t^2)$ and can be exacerbated by high control gains \cite{koch2019discrete}. Provided high sampling frequency and the adaptive law \eqref{eq:adptlaw} limiting the gains, the error due to discretization is negligible. This supports the validity of applying the continuous-time analysis to a discrete-time implementation.
\end{remark}

\begin{remark}
While Theorem 1 is presented in the context of a 7-DOF manipulator executing a 6-DOF task, the formulation is general to any manipulator with sufficient joint redundancy or to task spaces of lower dimension.
\end{remark}

\section{Simulation and Experimental Results} \label{sec:results}

\subsection{Implementation}
Simulations and experiments are performed using a Franka Emika 7-DOF manipulator, Gazebo simulation environment, and Robot Operating System (ROS). The controller and observer operate at the Franka Emika manipulator's real-time control interface frequency of $\Delta t=0.001$ s to satisfy the results in \cite{laila2002open} and ensure minimal discretization error \cite{koch2019discrete}. In simulation, the dynamic parameters $\mathcal{M}$, $\mathcal{C}$, $\mathcal{G}$, and $J$ are estimated using the penalty-based optimization method described in \cite{gaz2019dynamic}. In hardware experiments, the manipulator's internal software provides real-time numerical values for the dynamic parameters, which are used in the controller.

The proposed NT-STSM controller is compared with a proportional-derivative (PD) controller \cite{chen2023robotic}, NTSM controller \cite{feng2002non}, and STSM controller \cite{jeong2018tracking}, formulated as
\begin{align}
    \bm{u}_{\textrm{PD}} =& \ - K_p \bm{e} - 2 \sqrt{K_p} \dot{\bm{e}} + \bar{\mathcal{C}} + \bar{\mathcal{G}}, \nonumber \\
    \bm{u}_{\textrm{NT}} =& \ \bar{\mathcal{M}} \left( \ddot{\bm{x}}_{d} + H^{-1} \left( \frac{-\dot{\bm{e}}^{(2-\alpha)}}{\alpha \beta} - \kappa_1 \textrm{sign} (\bm{s}_1) \right) \right) + \bar{\mathcal{C}} + \bar{\mathcal{G}}, \nonumber \\
    \bm{u}_{\textrm{ST}} =& \ \bar{\mathcal{M}} \left( \ddot{\bm{x}}_{d} - H^{-1} \left( \kappa_1 |S_2|^{\frac{1}{2}} \textrm{sign} (\bm{s}_2) + \bm{\nu} \right) \right) + \bar{\mathcal{C}} + \bar{\mathcal{G}}, \nonumber  \\ \nonumber
    \dot{\bm{\nu}} =& \ - \kappa_2 \textrm{sign} (\bm{s}_2),  \nonumber
\end{align}
where $\bm{s}_1 = \ \bm{e} + \beta \dot{\bm{e}}^\alpha$ and $\bm{s}_2 = \ \bm{e} + \beta \dot{\bm{e}}$.

To smooth the control signal and mitigate the chattering in SM controllers, the $\textrm{sign}(\bm{s})$ terms are substituted with $\textrm{tanh}(k_s \bm{s})$, where $k_s$ is a positive design parameter that provides a trade-off between tracking accuracy and convergence speed. To ensure a fair comparison, all SM controllers were implemented using the same adaptive gain law \eqref{eq:adptlaw} with the same parameters, as shown in Table~\ref{tab:ctrlparam}. Three sets of PD controller gains were tested to determine a fair baseline to compare with the proposed controller, shown in Table~\ref{tab:ctrlparam}.


%
\begin{table}[h]
    \caption{Controller and observer parameters.}
    \centering 
    \begin{tabular}{c c c c}
        \hline
        PD-low & $K_{p \textrm{trans}} = 200$ & $K_{p \textrm{rot}} = 50$ & \\
        PD-med & $K_{p \textrm{trans}} = 800$ & $K_{p \textrm{rot}} = 200$ & \\
        PD-high & $K_{p \textrm{trans}} = 2000$ & $K_{p \textrm{rot}} = 500$ & \\
        \hline
        Sliding-Mode & $\beta = 1$ & $\alpha = 9/7$ & $k_s = 30$ \\
        Parameters& $\theta = 0.9$ & $\gamma = 6$ & $\Omega_1 = 1.5$ \\
        Eqs. \eqref{eqSS}, \eqref{eq:kappa} & $\Omega_2 = 0.14$ & & \\
        \hline 
        Adaptive & $\kappa_{1\textrm{min}} = 5$ & $\kappa_{1\textrm{max}} = 200$ & $\omega_\alpha = 1000$ \\ 
        Parameters & $\mu_{\alpha} = 0.001$ & $\eta_\alpha = 0.1$ &   \\ 
        Eq. \eqref{eq:adptlaw} & & & \\ \hline
        Observer & \multicolumn{3}{l}{$\bar{F} = \begin{bmatrix}
        20, & 20, & 20, & 20, & 20, & 20
        \end{bmatrix}^T$} \\
        Parameters & $\alpha_{o0} = 1.1\bar{F}$ & $\alpha_{o1} = 1.5 \bar{F}^{\frac{1}{2}}$ & $\alpha_{o2} = 1.9 \bar{F}^{\frac{1}{3}}$ \\
        Eqs. \eqref{eq:ob1}, \eqref{eq:ob2}, \eqref{eq:ob3} & $k_{o1} = 200$ & $k_{o2} = 400$ & $\alpha_e = 0.02$ \\ 
        & $\eta_q = 0.5$ &  & \\ \hline
    \end{tabular}
    \label{tab:ctrlparam}
\end{table}

To evaluate the robustness of the controllers to disturbances, the simulations are conducted with significant joint friction, unknown to the controllers. The added friction represents internal mechanical uncertainty, such as unmodeled joint friction from actuator degradation or wear. The friction parameters are detailed in Table~\ref{tab:fricparam}, where $c$ is the damping value, $\mu_s$ is the static friction value, $\mu_{k}$ is the Coulomb friction coefficient, and $\mu_{v}$ is the viscous friction coefficient for each joint. In the experiment, real-world disturbances, such as unmodeled joint friction, sensor noise, and dynamic model uncertainties, inherently test the robustness of the controllers.

\begin{figure}[ht]
    \centering
    \begin{subfigure}{0.44\columnwidth}
        \includegraphics[width=\linewidth]{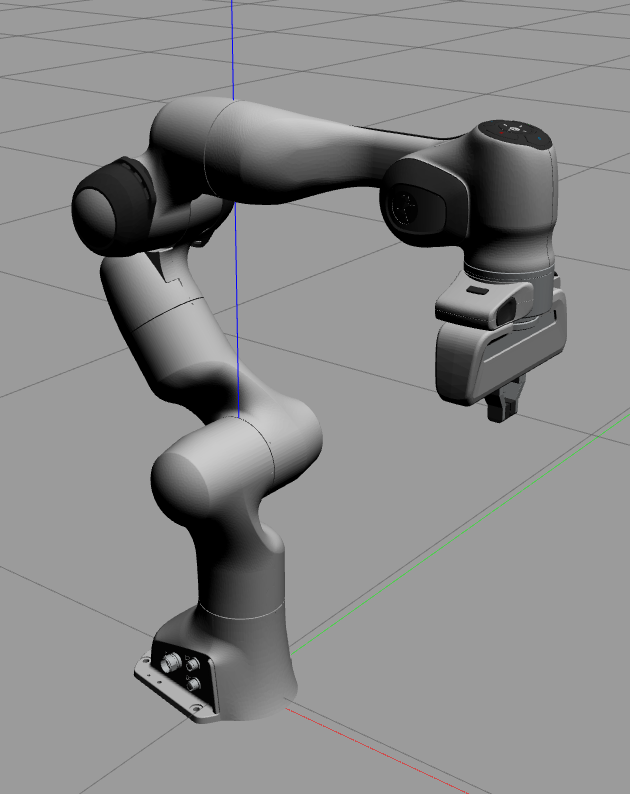}
        \caption{}
    \end{subfigure}
    \hspace{1mm}
    \begin{subfigure}{0.44\columnwidth}
        \includegraphics[width=\linewidth]{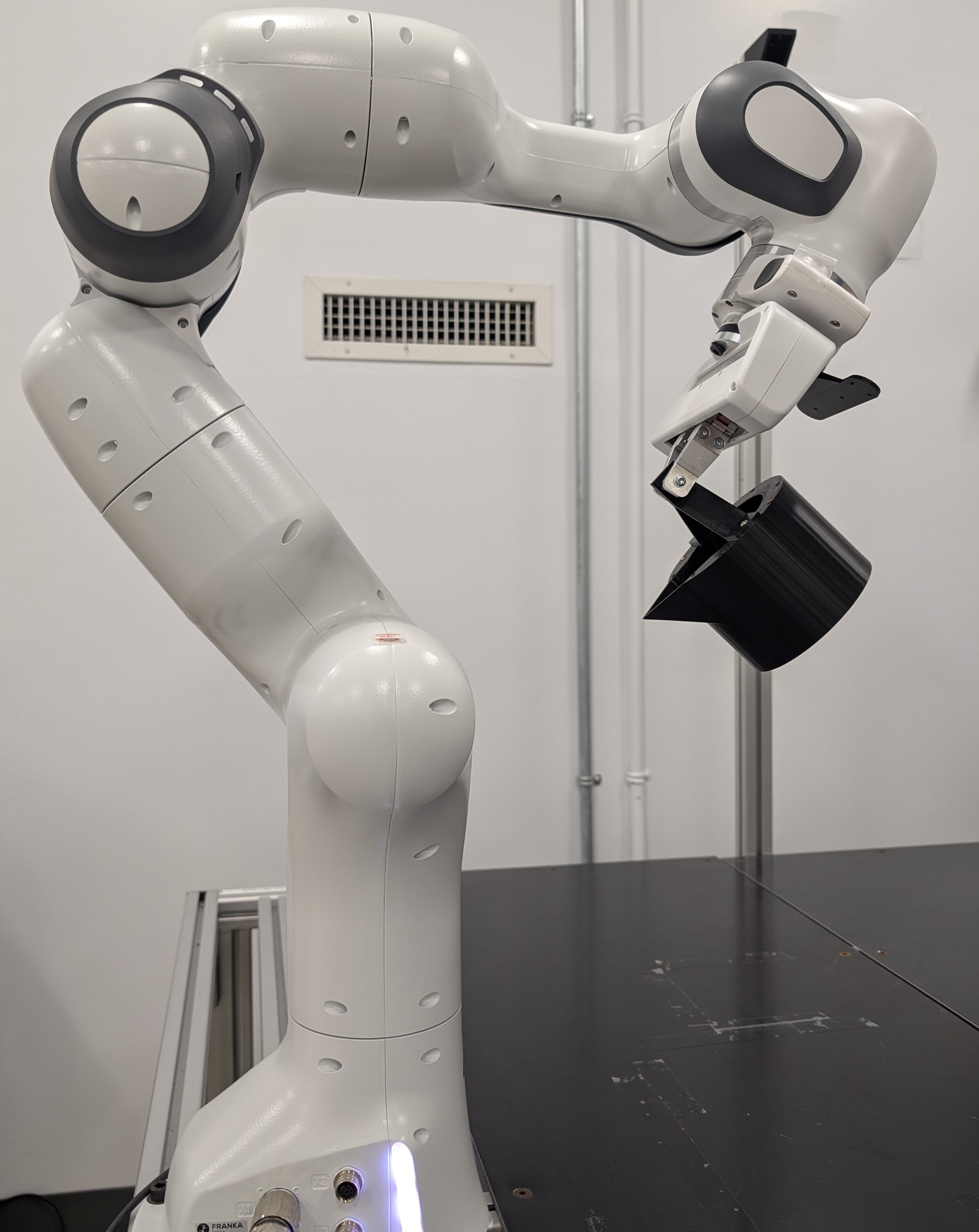}
        \caption{}
    \end{subfigure}

    \caption{Franka Emika 7-DOF manipulator. (a) Gazebo simulation environment; (b) Hardware setup grasping an object.}
    \label{fig:setup}
\end{figure}

\begin{table}[h]
    \caption{Simulation joint friction parameters.}
    \centering 
    \begin{tabular}{c c c c c}
        \hline
         & $c$ $(\textrm{Nms/rad})$ & $\mu_s$ $(\textrm{Nm})$ & $\mu_{k}$ & $\mu_{v}$\\ \hline
        Default & 0.003 & 0 & 0 & 16 \\ 
        Simulated & 0.003 & 0.5 & 25 & 25 \\ \hline
    \end{tabular}
    \label{tab:fricparam}
\end{table}

Fig.~\ref{fig:setup} shows the manipulator in the simulation and experimental setups. In the simulation, the robot executes a simultaneous motion of $0.05$ m translation in the $x$, $y$, and $z$ axes and $25^\circ$ rotation about the $x$, $y$, and $z$ axes. The controller is tested without external disturbances and then with external disturbances. The series of external disturbances applied to the end-effector is as follows. At $5$ s, $\bm{f}_{e}= \begin{bmatrix}
    0, & 5, & 0, & 0, & 0, & 0
\end{bmatrix}$, at $10$ s, $\bm{f}_{e}= \begin{bmatrix}
    5, & 0, & -5, & 0, & 0, & 0
\end{bmatrix}$, at $14$ s, $\bm{f}_{e}= \begin{bmatrix}
    0, & 0, & 0, & 1, & 0, & 0
\end{bmatrix}$, and at $17$ s, $\bm{f}_{e}= \begin{bmatrix}
    0, & -5, & 5, & 0, & -1, & 1
\end{bmatrix}$, where the first three elements of $\bm{f}_e$ have units of N and the last three have units of Nm. The disturbances are each applied for 1 s from each starting time. The disturbances are applied along different axes and at different times to test disturbance rejection performance across a variety of conditions. These forces and torques emulate environmental contacts, payload changes, or collisions. Their magnitudes are chosen to represent realistic interactions for the Franka Emika manipulator, considering the safety thresholds.

In the experiment, the robot manipulates a time-varying payload of a container holding steel balls with a total mass of 0.5 kg. The robot pours out the contents by executing a $0.1$ m translation in the $y$-axis and a  $40^\circ$ rotation about the $z$-axis. As the balls exit the container, the payload mass decreases nonlinearly, the center of mass shifts, and the rotational inertia changes. These variations result in unstructured, time-varying disturbances that affect translational and rotational dynamics.

The controller's tracking performance is quantified using the Root Mean Squared Error (RMSE). The translational tracking error, $\textrm{RMSE}_p$, is defined as the average RMSE over the three Cartesian axes:
\begin{equation*}
    \textrm{RMSE}_p = \frac{1}{3} \sum_{j=1}^{3} \sqrt{\frac{1}{N}  \sum_{i=1}^{N} \left( p_{j,i}-p_{d, j,i} \right) ^2}, 
\end{equation*}
where $N$ is the number of time steps and $j$ indexes the three axes. The rotational tracking error, $\textrm{RMSE}_{\xi}$, is calculated from the geodesic distance between the actual and desired unit quaternions using:
\begin{equation*}
     \textrm{RMSE}_{\xi} = \sqrt{\frac{4}{N} \sum_{i=1}^N \textrm{arccos}^2( \left| \bm{\xi} \cdot \bm{\xi}_d \right|)},
\end{equation*}
where $\xi_1 \cdot \xi_2$ denotes the quaternion dot product. To evaluate the performance along a specific rotational axis, the quaternion trajectories are converted to Euler angles, $\phi = [\phi_x, \phi_y, \phi_z]$, and the RMSE along a particular axis is given as:
\begin{equation*}
    \textrm{RMSE}_{\phi k} = \sqrt{\frac{1}{N}  \sum_{i=1}^{N} \left( \phi_{k,i}-\phi_{d, k,i}\right)^2},
\end{equation*}
where $k$ is the rotational axis of interest. To evaluate the chattering, the total variation of the control input is calculated as:
\begin{equation*}
    \textrm{TV}_{\tau} = \sum^7_{j=1} \sum^{N-1}_{i=1} |\tau_{j, i+1} - \tau_{j, i}|,
\end{equation*}
where a smaller value of $\textrm{TV}_{\tau}$ indicates a smoother control signal \cite{mondal2014adaptive}.

\subsection{Simulation Results}

\begin{figure}
    \centering
    \includegraphics[width=\linewidth]{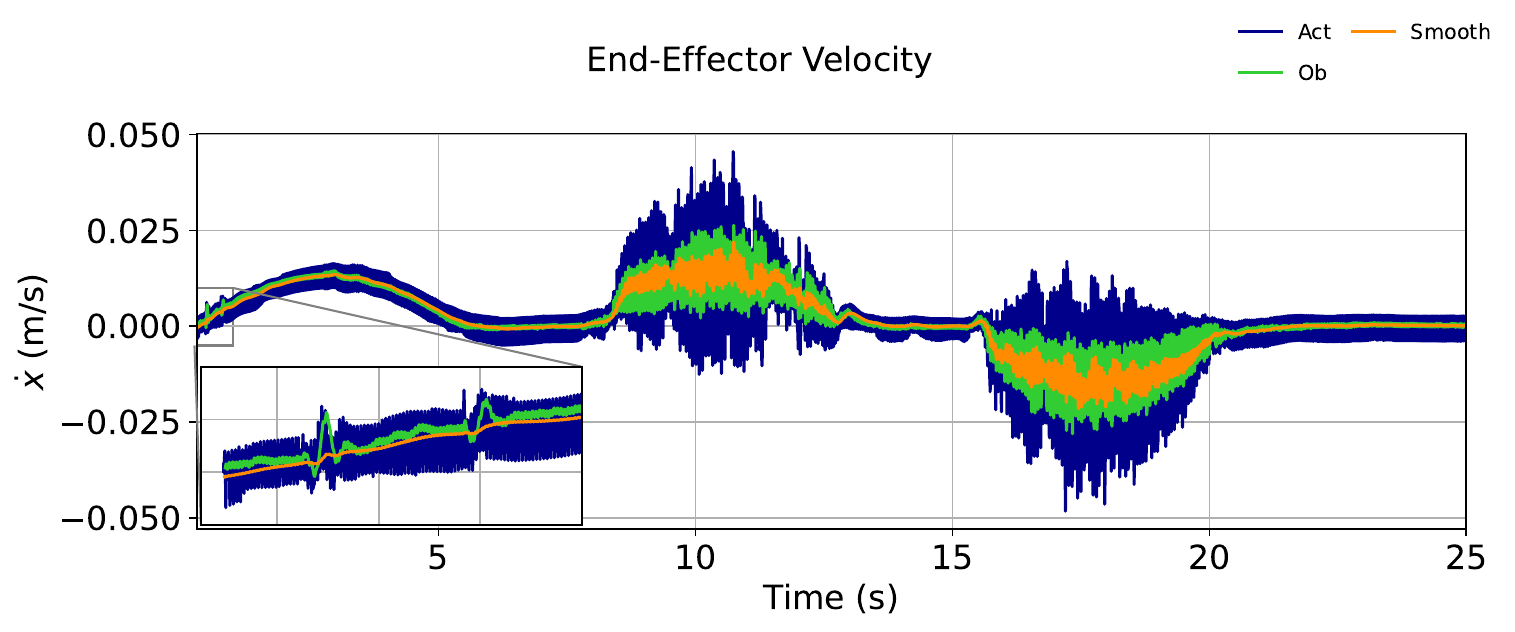}
    \caption{Comparison of x-axis velocity measurements.}
    \label{fig:sim0}
\end{figure}

Fig.~\ref{fig:sim0} compares different velocity estimation methods. The Franka robot's built-in velocity measurements (Act) are derived from position data, making them inherently noisy. Applying a low-pass filter (Smooth) reduces noise but introduces a measurement delay. During testing, both noise and delay in velocity estimates led to chattering in the task-space sliding-mode controllers. To mitigate this, the TOSM observer was implemented (Ob), where $\bar{F}$ is tuned based on the results in \cite{liu2022velocity} to provide a balance between accuracy, noise reduction, and minimal delay. To minimize the initial estimation error, the exponential moving average error is used with $\alpha_e =0.02$ to provide a smooth velocity estimate as the TOSM observer converges to the actual velocity.

Fig.~\ref{fig:sim0} also highlights why applying SM control is more challenging for rotational motion. From 0-5 s, when only translational motion occurs, the velocity noise is low. However, during rotational motion (9-13 s and 16-20 s), the noise is significantly higher, increasing susceptibility to chattering. 

\begin{figure}
    \centering
    \includegraphics[width=\linewidth]{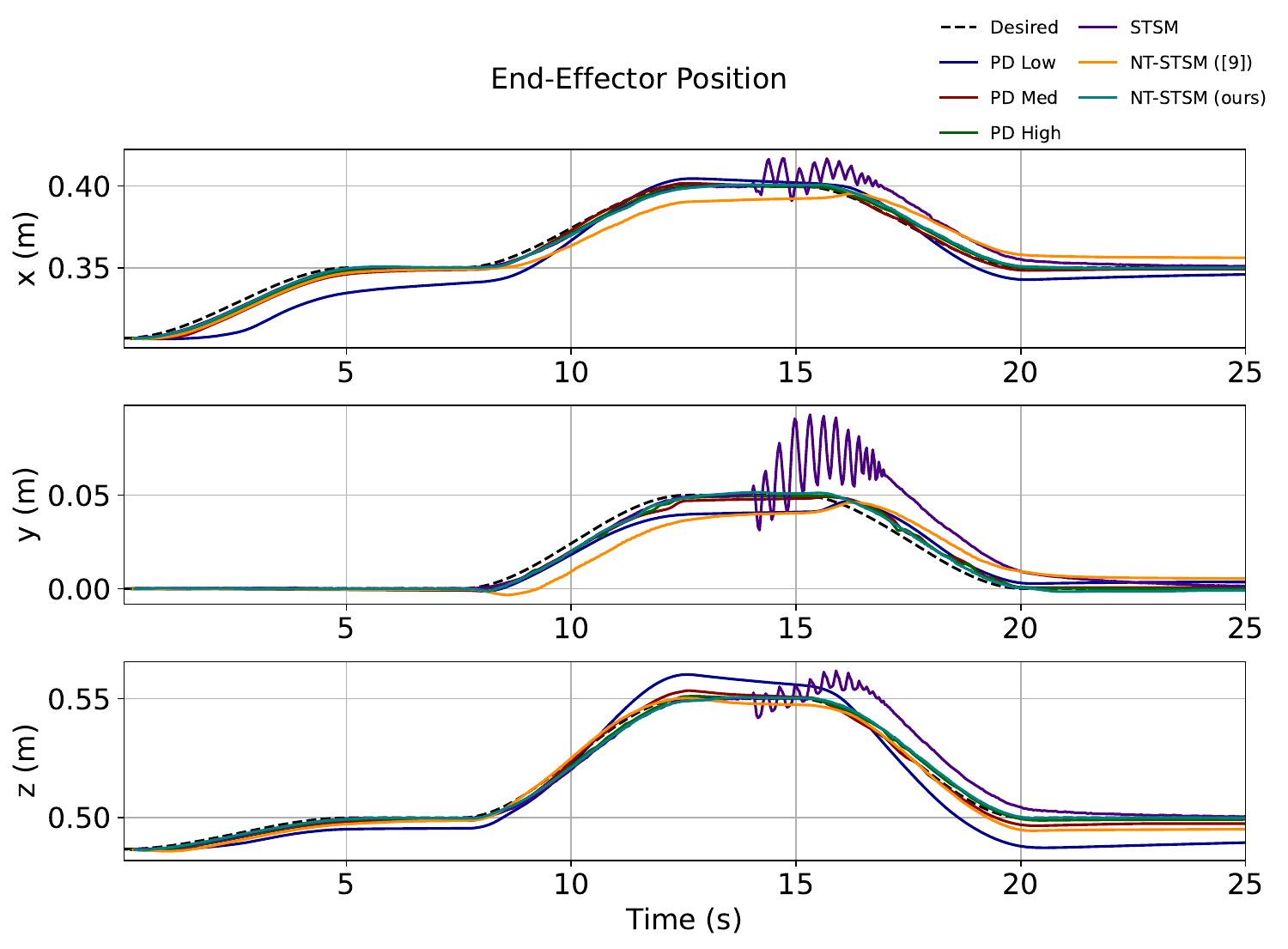}
    \caption{Simulation on the translational position tracking.}
    \label{fig:sim1}
\end{figure}

\begin{figure}
    \centering
    \includegraphics[width=\linewidth]{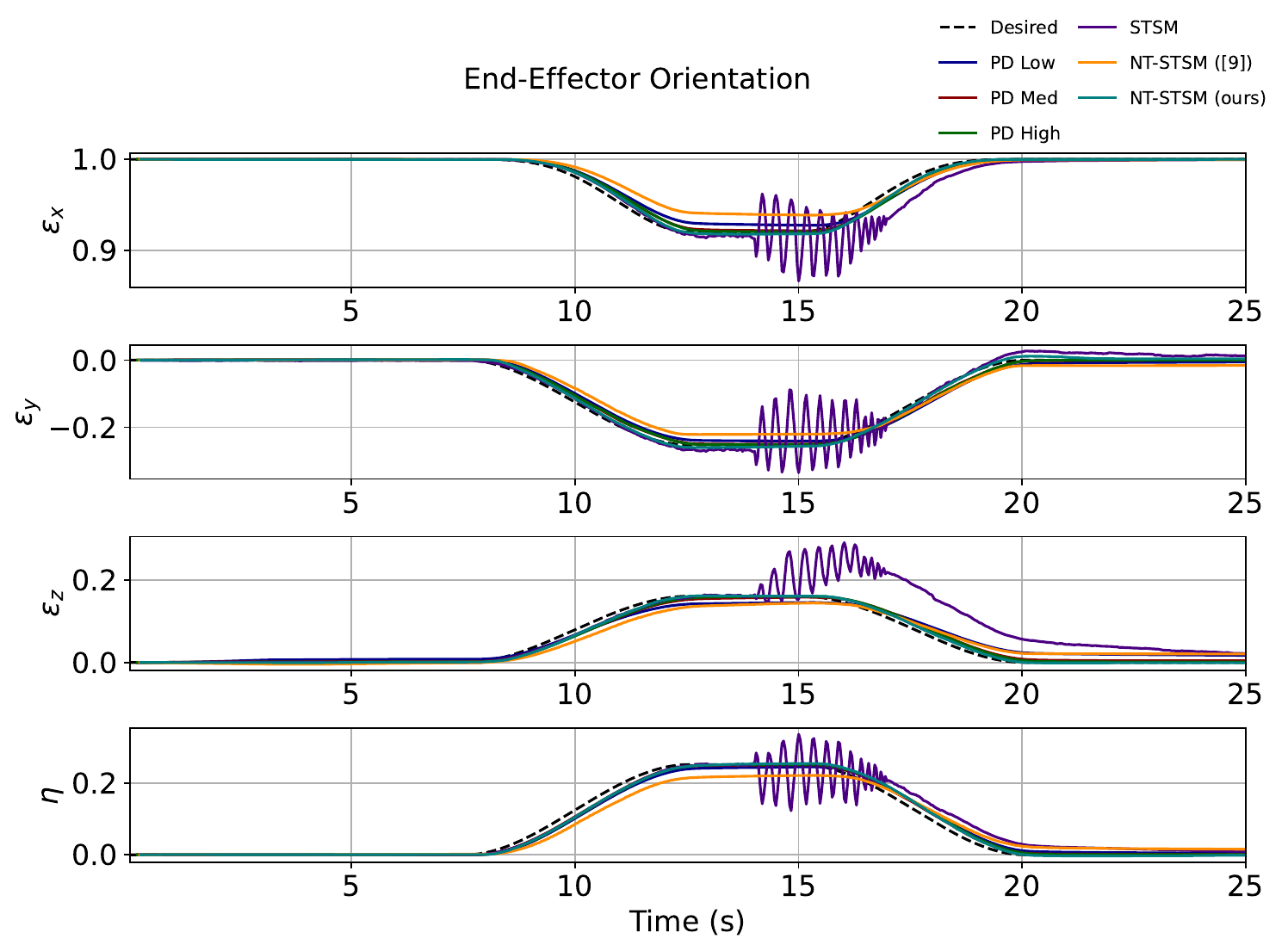}
    \caption{Simulation on the rotational position tracking.}
    \label{fig:sim2}
\end{figure}

Figs.~\ref{fig:sim1} and \ref{fig:sim2} show the position and orientation tracking results from the simulation without external disturbances. As expected, increasing the PD gains improved the tracking accuracy but increased the control effort, shown in Table~\ref{tab:avgerr}, where the performance metrics are Root Mean Squared Error (RMSE) and average control effort, $\tau_{\textrm{avg}}$. Note that the PD-low controller is significantly affected by the frictional disturbances, resulting in poor tracking.

Despite chattering reduction techniques, the NTSM controller exhibited severe chattering, leading to instability, and is omitted. The STSM controller performed well until 14 s, when it began to chatter significantly until stabilizing at 17 s. As shown in Table~\ref{tab:avgerr}, the proposed NT-STSM controller achieved comparable translational tracking to the PD-high controller and the highest rotational tracking accuracy of all the controllers while maintaining a low level of control effort. The NT-STSM controller also provides the lowest $\textrm{TV}_{\tau}$ among all controllers, indicating it has the smoothest control inputs. The NTSM and STSM controllers could be tuned to reduce chattering, however, this would require reducing $\kappa_1$ and $\kappa_2$, which would reduce the tracking accuracy and not be comparable to the accuracy of the NT-STSM controller.

Applying the NT-STSM controller from \cite{sun2024adaptive}, originally designed for the joint space, to the task space with its constrained relationship between $\kappa_1$ and $\kappa_2$, resulted in a 71.3\% increase in $\textrm{RMSE}_p$, and a 69.4\% increase in $\textrm{RMSE}_{\xi}$, compared to our approach. This result indicates that the strict gain relationship required by the stability proof in \cite{sun2024adaptive} limits the controller performance and is not suitable for this application. Compared to the PD-med controller, the proposed NT-STSM controller reduced $\textrm{RMSE}_p$ by 29.0\% and $\textrm{RMSE}_{\xi}$ by 27.6\%, with similar control effort. Compared to the PD-high controller, the proposed NT-STSM controller achieved similar tracking performance with 13.4\% lower control effort. While the PD-high controller achieves good tracking performance, it does so with the highest average control effort. This compromises energy efficiency and may exceed actuator limits. PD-med achieves a reasonable trade-off between accuracy and control effort, and is selected as the baseline in the following tasks.

\begin{table}[h]
	\caption{Performance metrics for the tracking task (simulation).}
    \centering 
    \begin{tabular}{c c c c c}
        \hline
        Controller & $\textrm{RMSE}_p$ & $\textrm{RMSE}_{\xi}$ & $\tau_{\textrm{avg}}$ & $\textrm{TV}_{\tau}$ \\
         & (m) & (rad) & (Nm) & (Nm) \\ \hline
        PD-low & 7.25e-3 & 4.91e-2 & \bf{3.69e-1} & 2.95e4 \\
        PD-med & 2.48e-3 & 3.11e-2 & 5.06e-1 & 6.02e4 \\
        PD-high & \bf{1.54e-3} & 2.50e-2 & 5.66e-1 & 8.67e4 \\
        NTSM & N/A & N/A & N/A & N/A \\
        STSM & 7.04e-3 & 1.19e-1 & 1.92 & 1.45e5 \\
        NT-STSM \cite{sun2024adaptive} & 6.13e-3 & 7.36e-2 & 5.10e-1 & 1.50e4 \\
        NT-STSM (ours) & 1.76e-3 & \bf{2.25e-2} & 4.90e-1 & \bf{8.45e3} \\ \hline
    \end{tabular}
    \label{tab:avgerr}
\end{table}

\begin{figure}
    \centering
    \includegraphics[width=\linewidth]{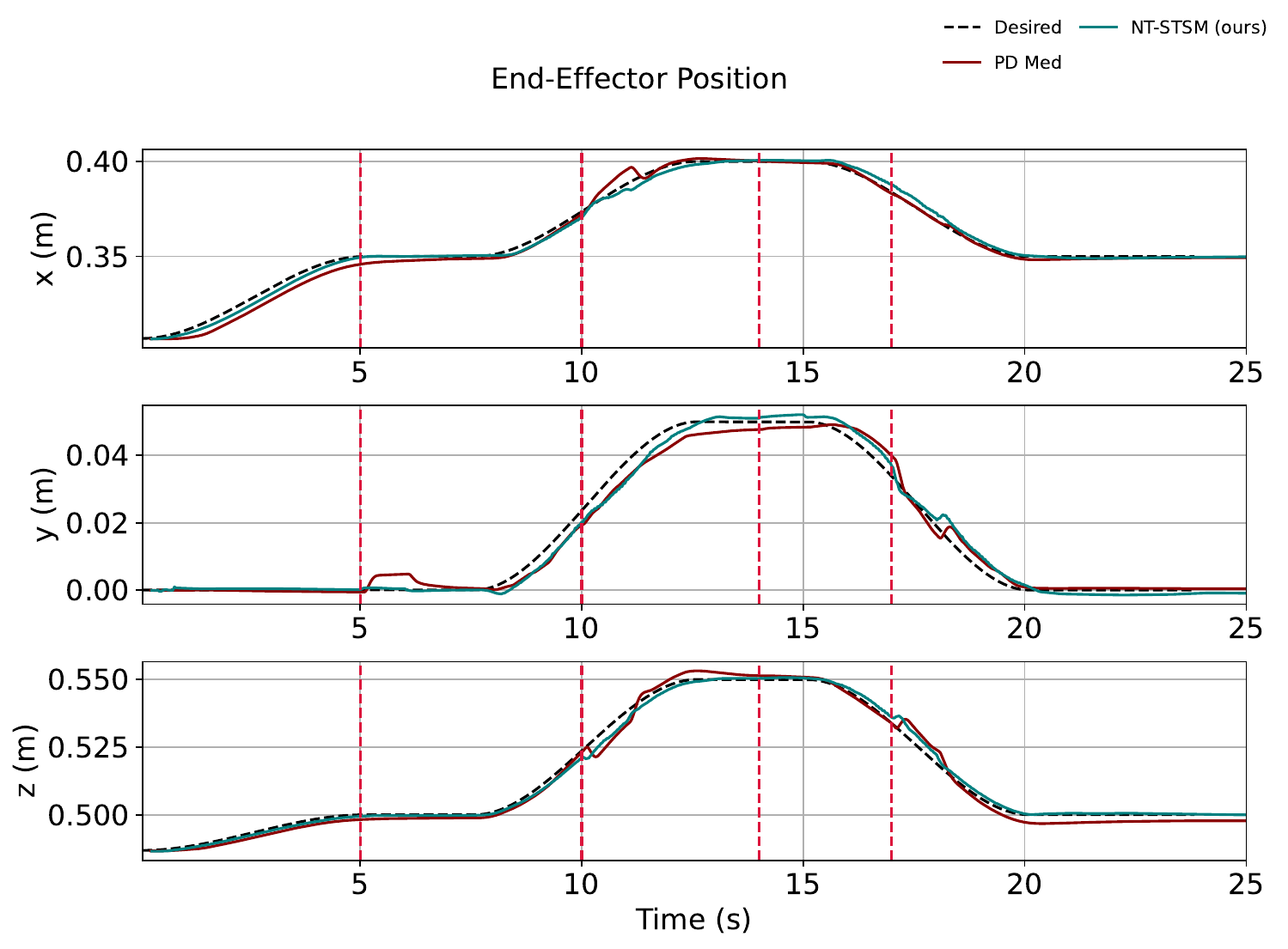}
    \caption{Simulation on the translational position tracking with external disturbances. The red dashed lines indicate when external forces are applied.}
    \label{fig:sim3}
\end{figure}

To further demonstrate the robustness of the NT-STSM controller, the same task was performed under a series of external forces and torques applied to the end-effector. Figure~\ref{fig:sim3} compares translational tracking performance, and Table~\ref{tab:avgerr2} shows the performance metrics, demonstrating that the PD-med controller is more affected by disturbances despite having similar control effort as the NT-STSM controller.

\begin{table}[h]
	\caption{Performance metrics for the tracking task with external disturbances (simulation).}
    \centering 
    \begin{tabular}{c c c c c}
        \hline
        Controller & $\textrm{RMSE}_p$ & $\textrm{RMSE}_{\xi}$ & $\tau_{\textrm{avg}}$ & $\textrm{TV}_{\tau}$ \\
         & (m) & (rad) & (Nm) & (Nm) \\ \hline
        PD-med & 2.72e-3 & 3.06e-2 & 7.13e-1 & 6.11e4 \\
        NT-STSM & \bf{1.89e-3} & \bf{2.80e-2} & 7.13e-1 & \bf{7.89e3} \\ \hline
    \end{tabular}
    \label{tab:avgerr2}
\end{table}

\subsection{Experimental Results}

\begin{figure}
    \centering
    \includegraphics[width=\linewidth]{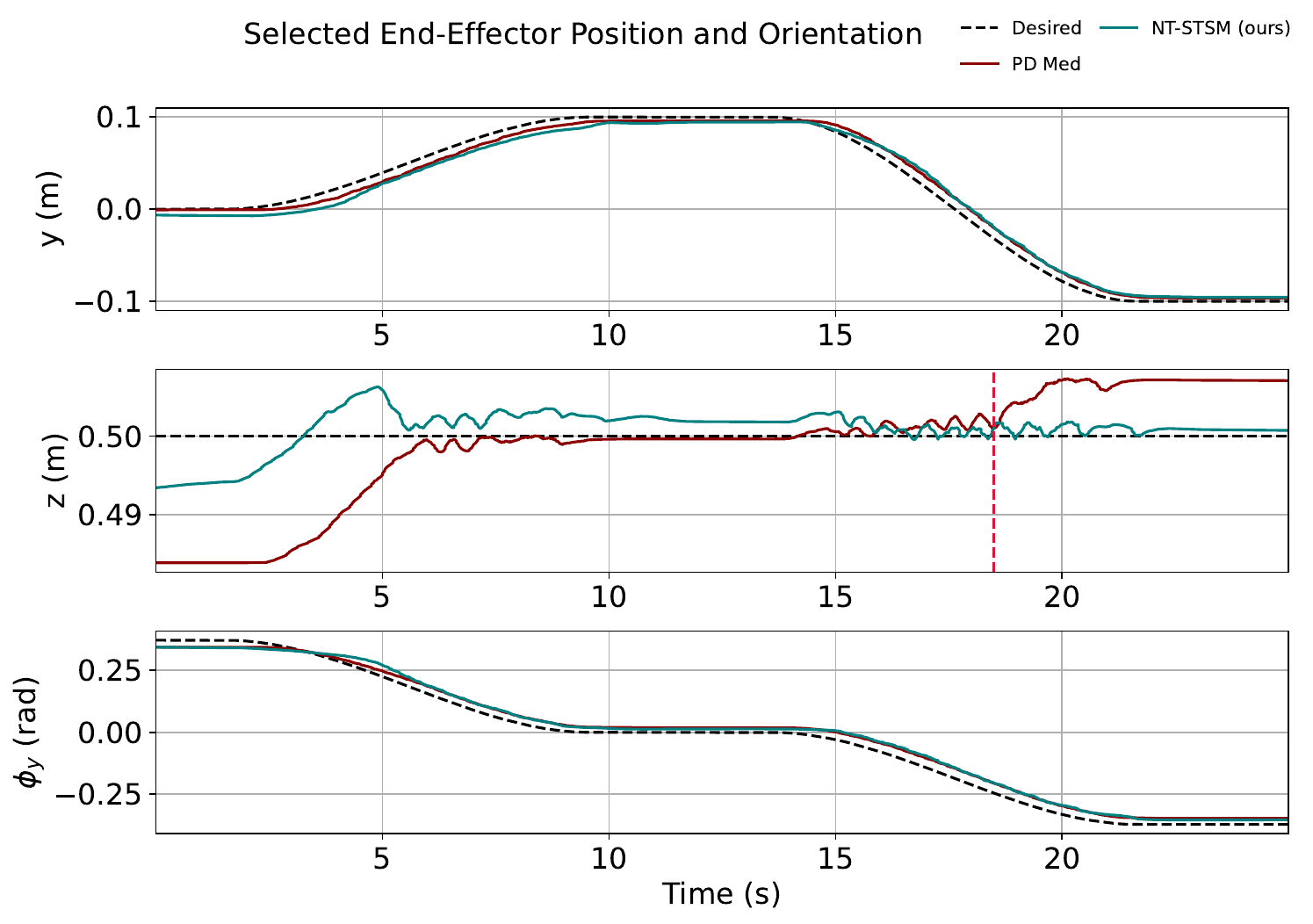}
    \caption{Experimental pouring task with selected position and orientation tracking. The payload starts changing at $t = 18$ s.}
    \label{fig:exp}
\end{figure}

A video of the experiment is available at \url{https://www.youtube.com/watch?v=ziweWAJxdYU}. Fig.~\ref{fig:exp} and Table \ref{tab:exp} compare the tracking performance of the PD-med and the NT-STSM controllers along the $y$ and $z$ translational axes and the $\epsilon_z$ rotational component. While the PD controller achieves lower tracking error in the $y$ direction and $\epsilon_z$ orientation, the NT-STSM controller demonstrates significantly improved tracking along the $z$-axis, which is most sensitive to the payload. This is clear at $t = 18$ s, where the payload change degrades the PD controller's performance while the NT-STSM controller maintains more accurate tracking. Additionally, the control input of the NT-STSM is significantly lower than the PD-med. These results confirm the robustness and efficiency of the proposed NT-STSM controller in the presence of unknown disturbances or varying payload and validate its viability for practical robotic applications involving external perturbations.

\begin{table}[h]
	\caption{Performance metrics for the pouring task (experiment).}
    \centering 
    \begin{tabular}{c c c c c c}
        \hline
        Controller & $\textrm{RMSE}_{py}$ & $\textrm{RMSE}_{pz}$ & $\textrm{RMSE}_{\phi y}$ & $\tau_{\textrm{avg}}$ & $\textrm{TV}_{\tau}$ \\
         & (m) & (m) & (rad) & (Nm) & (Nm) \\ \hline
        PD-med & \bf{7.30e-3} & 7.23e-3 & \bf{2.74e-2} & 1.30 & 2.07e4 \\
        NT-STSM & 9.41e-3 & \bf{2.74e-3} & 2.79e-2 & \bf{8.83e-1} & \bf{7.98e3} \\ \hline
    \end{tabular}
    \label{tab:exp}
\end{table}

\section{Conclusion} \label{sec:conclusion}
This paper proposed an adaptive NT-STSM controller formulated in task space for a 7-DOF robotic manipulator. A novel boundedness analysis established that tracking errors converge within a known region, providing flexible tuning guidelines. Compared to conventional controllers, the proposed method improves robustness to disturbances while reducing control effort and chattering, validating its applicability for real-time task space control of high-DOF manipulators.

The proposed controller achieves smooth, accurate behavior by combining $\tanh$ instead of $\mathrm{sign}$ to avoid discontinuities near the sliding surface, a super-twisting feedback control structure to smooth the control signal, and adaptive gains to reduce the control effort. The convergence proof provides flexible gain constraints that allow the controller to be tuned to improve tracking and reduce energy consumption compared to previous NT-STSM controllers. This work demonstrates that adaptive SM control in task space is an effective solution for precise robotic manipulation in environments with unmodeled disturbances and time-varying payloads. This work bridges the gap between theoretical robustness and practical implementation on real hardware and provides an avenue for applying SM control strategies in industrial and collaborative robotics.

Despite its advantages, one limitation of the proposed controller is that it requires manual tuning of multiple parameters, which requires dedicated operation time. Future work will focus on automating parameter selection through learning-based or optimization-based approaches. Based on the convincing results of this paper, future work will also include extending this controller to multi-manipulator systems and collaborative tasks by incorporating the force feedback.

%

%


\bibliographystyle{Bibliography/IEEEtranTIE}
\bibliography{Bibliography/BIB_xx-TIE-xxxx}\ 

\begin{thebibliography}{10}
\providecommand{\url}[1]{#1}
\csname url@samestyle\endcsname
\providecommand{\newblock}{\relax}
\providecommand{\bibinfo}[2]{#2}
\providecommand{\BIBentrySTDinterwordspacing}{\spaceskip=0pt\relax}
\providecommand{\BIBentryALTinterwordstretchfactor}{4}
\providecommand{\BIBentryALTinterwordspacing}{\spaceskip=\fontdimen2\font plus
\BIBentryALTinterwordstretchfactor\fontdimen3\font minus \fontdimen4\font\relax}
\providecommand{\BIBforeignlanguage}[2]{{%
\expandafter\ifx\csname l@#1\endcsname\relax
\typeout{** WARNING: IEEEtran.bst: No hyphenation pattern has been}%
\typeout{** loaded for the language `#1'. Using the pattern for}%
\typeout{** the default language instead.}%
\else
\language=\csname l@#1\endcsname
\fi
#2}}
\providecommand{\BIBdecl}{\relax}
\BIBdecl

\bibitem{feng2002non}
Y.~Feng, X.~Yu, and Z.~Man, ``Non-singular terminal sliding mode control of rigid manipulators,'' \emph{Automatica}, vol.~38, no.~12, pp. 2159--2167, 2002.

\bibitem{moreno2012strict}
J.~A. Moreno and M.~Osorio, ``Strict lyapunov functions for the super-twisting algorithm,'' \emph{IEEE Transactions on Automatic Control}, vol.~57, no.~4, pp. 1035--1040, 2012.

\bibitem{jeong2018tracking}
C.-S. Jeong, J.-S. Kim, and S.-I. Han, ``Tracking error constrained super-twisting sliding mode control for robotic systems,'' \emph{International Journal of Control, Automation and Systems}, vol.~16, pp. 804--814, 2018.

\bibitem{wang2016practical}
Y.~Wang, L.~Gu, Y.~Xu, and X.~Cao, ``Practical tracking control of robot manipulators with continuous fractional-order nonsingular terminal sliding mode,'' \emph{IEEE Transactions on Industrial Electronics}, vol.~63, no.~10, pp. 6194--6204, 2016.

\bibitem{sai2022approximate}
H.~Sai, Z.~Xu, C.~Xia, and X.~Sun, ``Approximate continuous fixed-time terminal sliding mode control with prescribed performance for uncertain robotic manipulators,'' \emph{Nonlinear Dynamics}, vol. 110, no.~1, pp. 431--448, 2022.

\bibitem{shtessel2012novel}
Y.~Shtessel, M.~Taleb, and F.~Plestan, ``A novel adaptive-gain supertwisting sliding mode controller: Methodology and application,'' \emph{Automatica}, vol.~48, no.~5, pp. 759--769, 2012.

\bibitem{gonzalez2011variable}
T.~Gonzalez, J.~A. Moreno, and L.~Fridman, ``Variable gain super-twisting sliding mode control,'' \emph{IEEE Transactions on Automatic Control}, vol.~57, no.~8, pp. 2100--2105, 2011.

\bibitem{baek2016new}
J.~Baek, M.~Jin, and S.~Han, ``A new adaptive sliding-mode control scheme for application to robot manipulators,'' \emph{IEEE Transactions on Industrial Electronics}, vol.~63, no.~6, pp. 3628--3637, 2016.

\bibitem{zhang2021adaptive}
L.~Zhang, H.~Liu, D.~Tang, Y.~Hou, and Y.~Wang, ``Adaptive fixed-time fault-tolerant tracking control and its application for robot manipulators,'' \emph{IEEE Transactions on Industrial Electronics}, vol.~69, no.~3, pp. 2956--2966, 2021.

\bibitem{sun2022adaptive}
Y.~Sun, J.~Liu, Y.~Gao, Z.~Liu, and Y.~Zhao, ``Adaptive neural tracking control for manipulators with prescribed performance under input saturation,'' \emph{IEEE/ASME Transactions on Mechatronics}, vol.~28, no.~2, pp. 1037--1046, 2022.

\bibitem{hu2022robust}
Y.~Hu, H.~Yan, H.~Zhang, M.~Wang, and L.~Zeng, ``Robust adaptive fixed-time sliding-mode control for uncertain robotic systems with input saturation,'' \emph{IEEE Transactions on Cybernetics}, vol.~53, no.~4, pp. 2636--2646, 2022.

\bibitem{jie2020trajectory}
W.~Jie, Z.~Yudong, B.~Yulong, H.~H. Kim, and M.~C. Lee, ``Trajectory tracking control using fractional-order terminal sliding mode control with sliding perturbation observer for a 7-dof robot manipulator,'' \emph{IEEE/ASME Transactions on Mechatronics}, vol.~25, no.~4, pp. 1886--1893, 2020.

\bibitem{veysi2015novel}
M.~Veysi, M.~R. Soltanpour, and M.~H. Khooban, ``A novel self-adaptive modified bat fuzzy sliding mode control of robot manipulator in presence of uncertainties in task space,'' \emph{Robotica}, vol.~33, no.~10, pp. 2045--2064, 2015.

\bibitem{lewis2003robot}
F.~L. Lewis, D.~M. Dawson, and C.~T. Abdallah, \emph{Robot manipulator control: theory and practice}.\hskip 1em plus 0.5em minus 0.4em\relax CRC Press, 2003.

\bibitem{khan2011task}
S.~G. Khan, J.~Jalani, G.~Herrmann, T.~Pipe, and C.~Melhuish, ``Task space integral sliding mode controller implementation for 4{DOF} of a humanoid bert ii arm with posture control,'' in \emph{Towards Autonomous Robotic Systems: 12th Annual Conference, TAROS 2011, Sheffield, UK, August 31--September 2, 2011. Proceedings 12}, pp. 299--310.\hskip 1em plus 0.5em minus 0.4em\relax Springer, 2011.

\bibitem{nicolis2020operational}
D.~Nicolis, F.~Allevi, and P.~Rocco, ``Operational space model predictive sliding mode control for redundant manipulators,'' \emph{IEEE Transactions on Robotics}, vol.~36, no.~4, pp. 1348--1355, 2020.

\bibitem{fateh2024model}
A.~Fateh and H.~Momeni, ``Model-free adaptive task-space sliding mode control of a delta robot using a novel reaching law,'' \emph{ISA Transactions}, vol. 149, pp. 69--80, 2024.

\bibitem{yi2019adaptive}
S.~Yi and J.~Zhai, ``Adaptive second-order fast nonsingular terminal sliding mode control for robotic manipulators,'' \emph{ISA Transactions}, vol.~90, pp. 41--51, 2019.

\bibitem{zhai2021fast}
J.~Zhai and Z.~Li, ``Fast-exponential sliding mode control of robotic manipulator with super-twisting method,'' \emph{IEEE Transactions on Circuits and Systems II: Express Briefs}, vol.~69, no.~2, pp. 489--493, 2021.

\bibitem{cruz2022non}
D.~Cruz-Ortiz, I.~Chairez, and A.~Poznyak, ``Non-singular terminal sliding-mode control for a manipulator robot using a barrier lyapunov function,'' \emph{ISA Transactions}, vol. 121, pp. 268--283, 2022.

\bibitem{sun2024adaptive}
C.~Sun, Z.~Huang, and H.~Wu, ``Adaptive super-twisting global nonsingular terminal sliding mode control for robotic manipulators,'' \emph{Nonlinear Dynamics}, pp. 1--11, 2024.

\bibitem{siciliano2008springer}
B.~Siciliano, O.~Khatib, and T.~Kr{\"o}ger, \emph{Springer handbook of robotics}, vol. 200.\hskip 1em plus 0.5em minus 0.4em\relax Springer, 2008.

\bibitem{mukundan2012advanced}
R.~Mukundan, \emph{Advanced methods in computer graphics: with examples in OpenGL}.\hskip 1em plus 0.5em minus 0.4em\relax Springer Science \& Business Media, 2012.

\bibitem{kreyszig2008advanced}
E.~Kreyszig, K.~Stroud, and G.~Stephenson, ``Advanced engineering mathematics,'' \emph{Integration}, vol.~9, no.~4, 2008.

\bibitem{kou2018linear}
K.~I. Kou and Y.-H. Xia, ``Linear quaternion differential equations: basic theory and fundamental results,'' \emph{Studies in Applied Mathematics}, vol. 141, no.~1, pp. 3--45, 2018.

\bibitem{liu2022velocity}
S.~B. Liu, A.~Giusti, and M.~Althoff, ``Velocity estimation of robot manipulators: An experimental comparison,'' \emph{IEEE Open Journal of Control Systems}, vol.~2, pp. 1--11, 2022.

\bibitem{van2013output}
M.~Van, H.-J. Kang, Y.-S. Suh, and K.-S. Shin, ``Output feedback tracking control of uncertain robot manipulators via higher-order sliding-mode observer and fuzzy compensator,'' \emph{Journal of Mechanical Science and Technology}, vol.~27, pp. 2487--2496, 2013.

\bibitem{siciliano2008robotics}
B.~Siciliano, L.~Sciavicco, L.~Villani, and G.~Oriolo, \emph{Robotics: Modelling, Planning and Control}, 1st~ed.\hskip 1em plus 0.5em minus 0.4em\relax Springer Publishing Company, Incorporated, 2008.

\bibitem{campa2006kinematic}
R.~Campa, K.~Camarillo, and L.~Arias, ``Kinematic modeling and control of robot manipulators via unit quaternions: Application to a spherical wrist,'' in \emph{Proceedings of the 45th IEEE Conference on Decision and Control}, pp. 6474--6479.\hskip 1em plus 0.5em minus 0.4em\relax IEEE, 2006.

\bibitem{bhat2000finite}
S.~P. Bhat and D.~S. Bernstein, ``Finite-time stability of continuous autonomous systems,'' \emph{SIAM Journal on Control and Optimization}, vol.~38, no.~3, pp. 751--766, 2000.

\bibitem{moreno2008lyapunov}
J.~A. Moreno and M.~Osorio, ``A lyapunov approach to second-order sliding mode controllers and observers,'' in \emph{2008 47th IEEE Conference on Decision and Control}, pp. 2856--2861.\hskip 1em plus 0.5em minus 0.4em\relax IEEE, 2008.

\bibitem{poznyak2009advanced}
A.~S. Poznyak, \emph{Advanced mathematical tools for automatic control engineers: Stochastic techniques}.\hskip 1em plus 0.5em minus 0.4em\relax Elsevier Ltd, 2009.

\bibitem{Lemma_ref}
F.~Plestan, Y.~Shtessel, V.~Bregeault, and A.~Poznyak, ``New methodologies for adaptive sliding mode control,'' \emph{International Journal of Control}, vol.~83, no.~9, pp. 1907--1919, 2010.

\bibitem{smith2024adaptive}
S.~Smith and Y.-J. Pan, ``Adaptive observer-based super-twisting sliding mode control for low altitude quadcopter grasping,'' \emph{IEEE/ASME Transactions on Mechatronics}, 2024.

\bibitem{slotine1991nonlinear}
J.-J.~E. Slotine and W.~Li, \emph{Applied Nonlinear Control}.\hskip 1em plus 0.5em minus 0.4em\relax Prentice Hall, 1991.

\bibitem{laila2002open}
D.~S. Laila, D.~Ne{\v{s}}i{\'c}, and A.~R. Teel, ``Open-and closed-loop dissipation inequalities under sampling and controller emulation,'' \emph{European Journal of Control}, vol.~8, no.~2, pp. 109--125, 2002.

\bibitem{koch2019discrete}
S.~Koch and M.~Reichhartinger, ``Discrete-time equivalents of the super-twisting algorithm,'' \emph{Automatica}, vol. 107, pp. 190--199, 2019.

\bibitem{gaz2019dynamic}
C.~Gaz, M.~Cognetti, A.~Oliva, P.~R. Giordano, and A.~De~Luca, ``Dynamic identification of the franka emika panda robot with retrieval of feasible parameters using penalty-based optimization,'' \emph{IEEE Robotics and Automation Letters}, vol.~4, no.~4, pp. 4147--4154, 2019.

\bibitem{chen2023robotic}
Q.~Chen, L.~Wan, and Y.-J. Pan, ``Robotic pick-and-handover maneuvers with camera-based intelligent object detection and impedance control,'' \emph{Transactions of the Canadian Society for Mechanical Engineering}, vol.~47, no.~4, pp. 486--496, 2023.

\bibitem{mondal2014adaptive}
S.~Mondal and C.~Mahanta, ``Adaptive second order terminal sliding mode controller for robotic manipulators,'' \emph{Journal of the Franklin Institute}, vol. 351, no.~4, pp. 2356--2377, 2014.

\end{thebibliography}

\vspace{-1cm}
\begin{IEEEbiography}[{\includegraphics[width=1in,height=1.25in,clip,keepaspectratio]{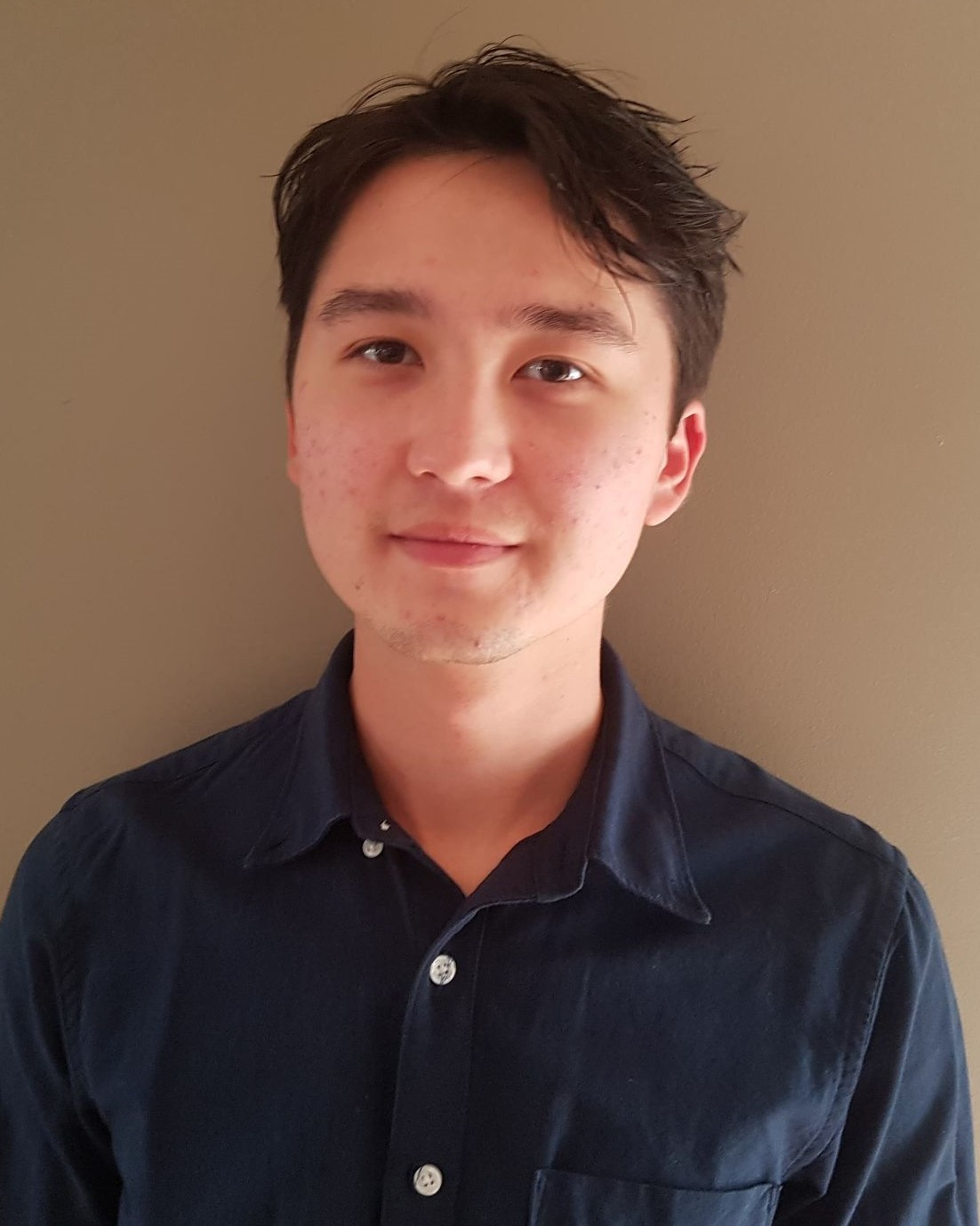}}]
{Lucas Wan} received the B.E. and M.ASc. degrees in mechanical engineering from Dalhousie University, Halifax, NS, Canada, in 2019 and 2021, respectively, where he is currently pursuing the Ph.D. degree in mechanical engineering. His research interests include multiagent systems, collaborative robotics, and reinforcement learning.
\end{IEEEbiography}

\begin{IEEEbiography}[{\includegraphics[width=1in,height=1.25in,clip,keepaspectratio]{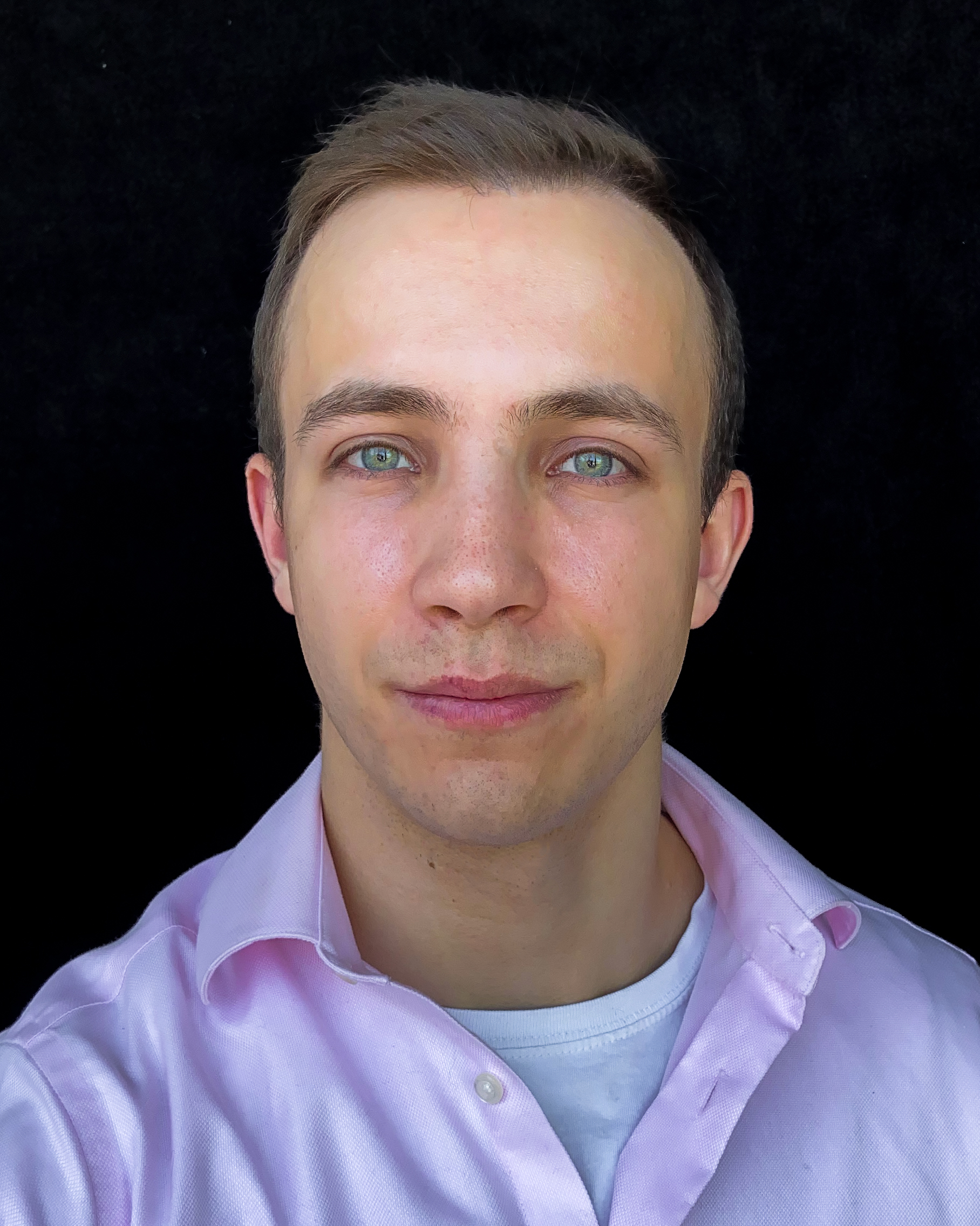}}]
{Sean Smith} received the B.E. and M.ASc. degrees in mechanical engineering from Dalhousie University, Halifax, NS, Canada, in 2021 and 2023, respectively. He is currently working towards his Ph.D. in mechanical engineering from the Universit\'e Grenoble Alpes, France. His research interests include nonlinear systems, robust control, optimal control, and robotics.
\end{IEEEbiography}

\begin{IEEEbiography}[{\includegraphics[width=1in,height=1.25in,clip,keepaspectratio]{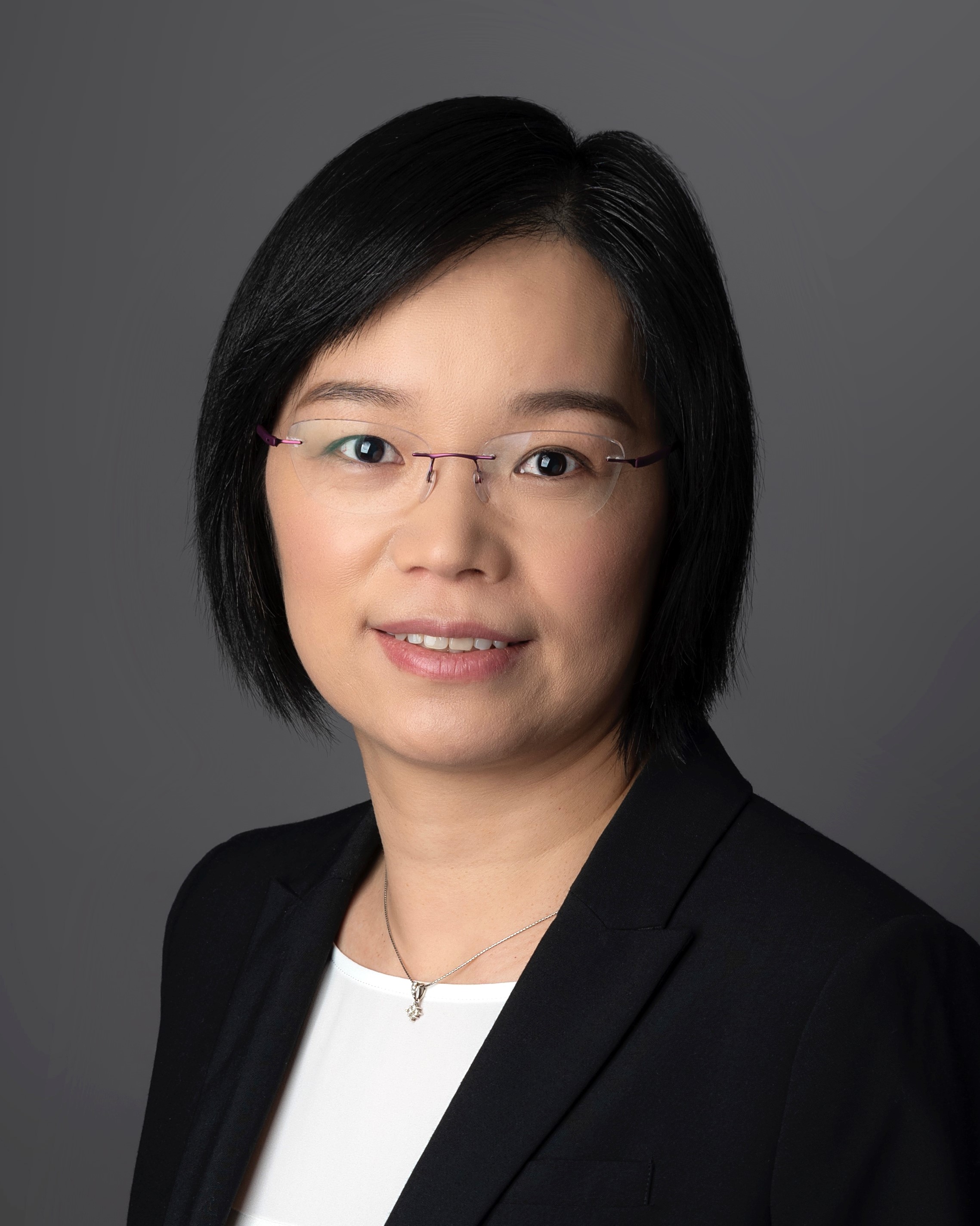}}]
{Ya-Jun Pan} is a Professor in the Dept. of Mechanical Engineering at Dalhousie University, Canada. She received the B.E. degree in Mechanical Engineering from Yanshan University, the M.E. degree in Mechanical Engineering from Zhejiang University, and the Ph.D. degree in Electrical and Computer Engineering from the National University of Singapore. 

She held post-doctoral positions of CNRS in the Laboratoire d'Automatique de Grenoble in France and the Dept. of Electrical and Computer Engineering at the University of Alberta in Canada. Her research interests are robust nonlinear control, cyber physical systems, intelligent transportation systems, haptics, and collaborative multiple robotic systems. She is a Fellow of Canadian Academy of Engineering (CAE), Engineering Institute of Canada (EIC), ASME, CSME, a Senior member of IEEE, and a registered Professional Engineer in Nova Scotia, Canada.
\end{IEEEbiography}

\begin{IEEEbiography}[{\includegraphics[width=1in,height=1.25in,clip,keepaspectratio]{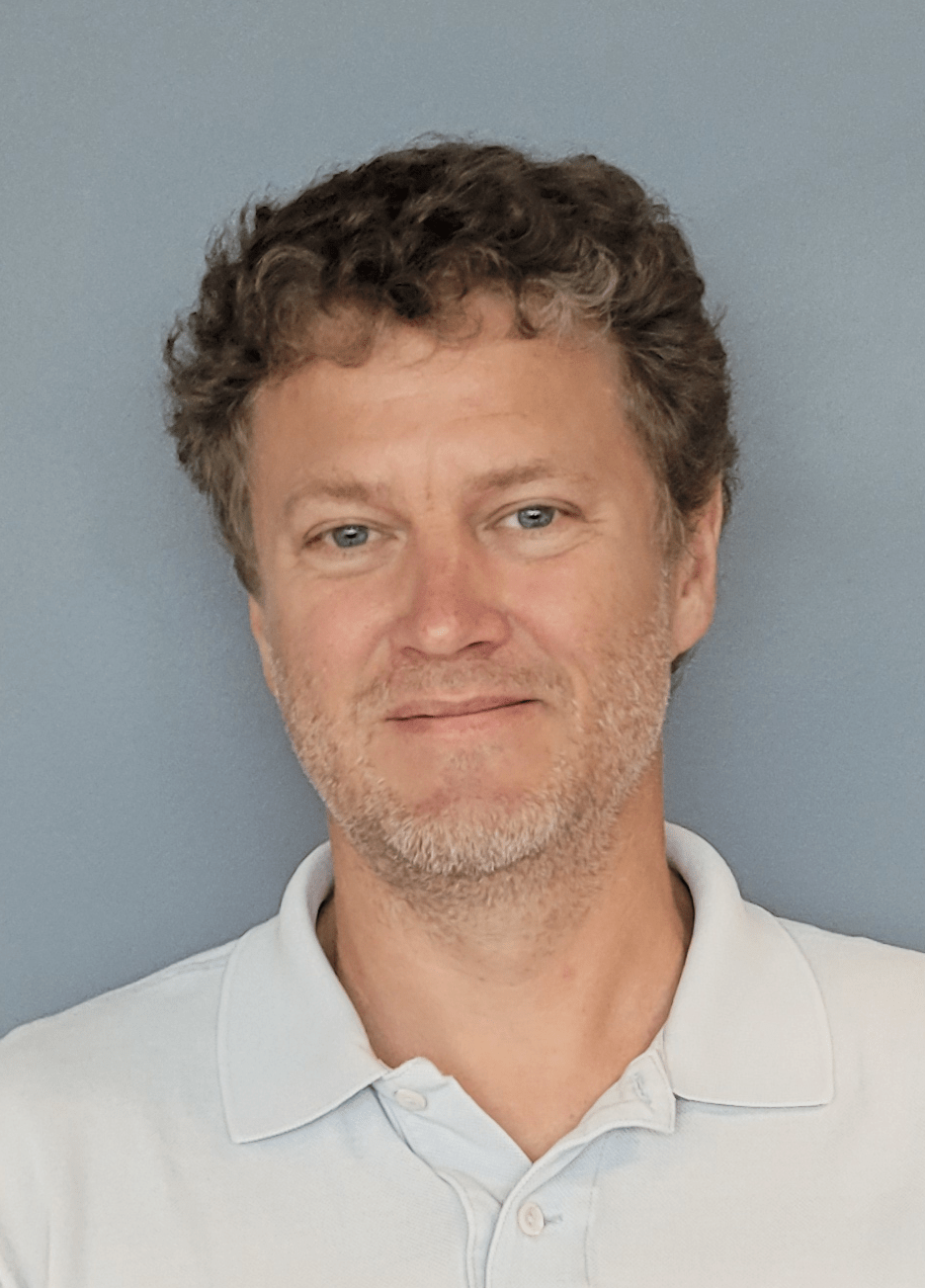}}]
{Emmanuel Witrant} obtained a B.Sc. in Aerospace Eng. from Georgia Tech in 2001 and a Ph.D. in Automatic Control from Grenoble University in 2005. He joined the Physics department of Univ. Grenoble Alpes and GIPSA-lab as an Associate Professor in 2007 and became Professor in 2020.

His research interest is focused on finding new methods for modeling and control of inhomogeneous transport phenomena (of information, energy, gases...), with real-time and/or optimization constraints. Such methods provide new results for automatic control (time-delay and distributed systems), controlled thermonuclear fusion (temperature/magnetic flux profiles in tokamak plasma), environmental sciences (atmospheric history of trace gas from ice cores measurements) and Poiseuille's flows (ventilation control in mines, car engines and intelligent buildings).
\end{IEEEbiography}

\end{document}